\documentclass[aps,pra,superscriptaddress,reprint]{revtex4-1}

\usepackage{graphicx}
\usepackage{dcolumn}
\usepackage{bm}
\usepackage[normalem]{ulem}	
\usepackage{xcolor}

\usepackage{ifthen}
\usepackage[mathscr]{euscript}
\usepackage{mathtools}
\usepackage{amsmath,amssymb,amsthm}
\usepackage{bbold}

\DeclareMathOperator{\tr}{tr}

\DeclareMathOperator{\dham}{d_{H}}
\DeclareMathOperator{\wham}{w_{H}}

\newcommand{\ket}[1]{\vert#1\rangle}
\newcommand{\bra}[1]{\langle#1\vert}
\newcommand{\e}[0]{\mathrm{e}}

\begin{document}

\title{Experimental bit commitment based on quantum communication and special relativity}

\author{T.~Lunghi}
\affiliation{Group of Applied Physics, University of Geneva, CH-1211 Gen\`eve 4, Switzerland}
\author{J.~Kaniewski}
\affiliation{Centre for Quantum Technologies, National University of Singapore, 3 Science Drive 2, Singapore 117543}
\author{F.~Bussi\`{e}res}
\affiliation{Group of Applied Physics, University of Geneva, CH-1211 Gen\`eve 4, Switzerland}
\author{R.~Houlmann}
\affiliation{Group of Applied Physics, University of Geneva, CH-1211 Gen\`eve 4, Switzerland}
\author{M.~Tomamichel}
\affiliation{Centre for Quantum Technologies, National University of Singapore, 3 Science Drive 2, Singapore 117543}
\author{A.~Kent}
\affiliation{Centre for Quantum Information and Foundations, DAMTP, Centre for Mathematical Sciences, University of Cambridge, United Kingdom}
\affiliation{Perimeter Institute for Theoretical Physics, 31 Caroline Street North, Waterloo, Ontario, Canada N2L 2Y5}
\author{N.~Gisin}
\affiliation{Group of Applied Physics, University of Geneva, CH-1211 Gen\`eve 4, Switzerland}
\author{S.~Wehner}
\affiliation{Centre for Quantum Technologies, National University of Singapore, 3 Science Drive 2, Singapore 117543}
\author{H.~Zbinden}
\affiliation{Group of Applied Physics, University of Geneva, CH-1211 Gen\`eve 4, Switzerland}
\date{\today}

\begin{abstract}
  Bit commitment is a fundamental cryptographic primitive in which Bob
  wishes to commit a secret bit to Alice. Perfectly secure bit
  commitment is impossible through asynchronous exchange of quantum
  information. Perfect security is however possible when Alice and Bob
  split into several agents exchanging classical and quantum
  communication at times and locations suitably chosen to satisfy
  specific relativistic constraints. Here we report an implementation
  of a bit commitment protocol using quantum communication and special
  relativity. Our protocol is based on [A.~Kent, Phys. Rev. Lett.
  \textbf{109}, 130501 (2012)] and has the advantage that it is
  practically feasible with arbitrary large separations between the
  agents in order to maximize the commitment time. By positioning
  agents in Geneva and Singapore, we obtained a commitment time of
  15~ms. A security analysis considering experimental imperfections
  and finite statistics is presented.
\end{abstract}

\maketitle

Bit commitment is a fundamental primitive with several applications
such as coin tossing~\cite{Blum1981a} and secure
voting~\cite{Broadbent2008a}. In a bit commitment protocol, Bob
commits a secret bit to Alice at a given instant which he can choose
to reveal some time later. Security here means that if Bob is honest,
then his bit is perfectly concealed from Alice until he decides to
open the commitment and reveal his bit. Furthermore, if Alice is
honest, then it should be impossible for Bob to change his mind once
the commitment is made. That is, the only bit he can reveal is the one
he originally committed himself to.  Information-theoretically secure
bit commitment in a setting where the two parties exchange classical
messages in an asynchronous fashion is impossible.  An extensive
amount of work was devoted to study asynchronous
quantum bit commitment, for which perfect security was ultimately shown to be
impossible~\cite{Mayers1997a,Lo1997a} as well (see
also~\cite{Dariano2007a,winkler11}).
This does not preclude the existence of protocols with partial (but less-than-complete) bias~\cite{Aharonov2000a,Spekkens2001a,Kitaev2003a,Chailloux2011a}, which have been the subject of related experimental work~\cite{Nguyen2008a,Berlin2011a,Pappa2013a}. Bit commitment was also demonstrated experimentally using the assumption of noisy quantum storage~\cite{Ng2012a}. Alternatively, perfectly secure relativistic protocols based on the exchange of quantum and classical
bits have been proposed~\cite{Kent1999, Kent2005, Kent2011, Kent2012a}.   We focus here on the
protocol~\cite{Kent2012a}, which is proven secure in the ideal case
(i.e.~with perfect devices)~\cite{Croke2012a,Kaniewski2013a} and in
the presence of losses~\cite{Croke2012a}. In this
Letter, we present the first experimental implementation of a secure
bit commitment protocol that is based on quantum communication and
relativistic constraints, along with a security proof taking into
account its experimental imperfections.

\begin{figure}[!t]
\includegraphics[width=8cm]{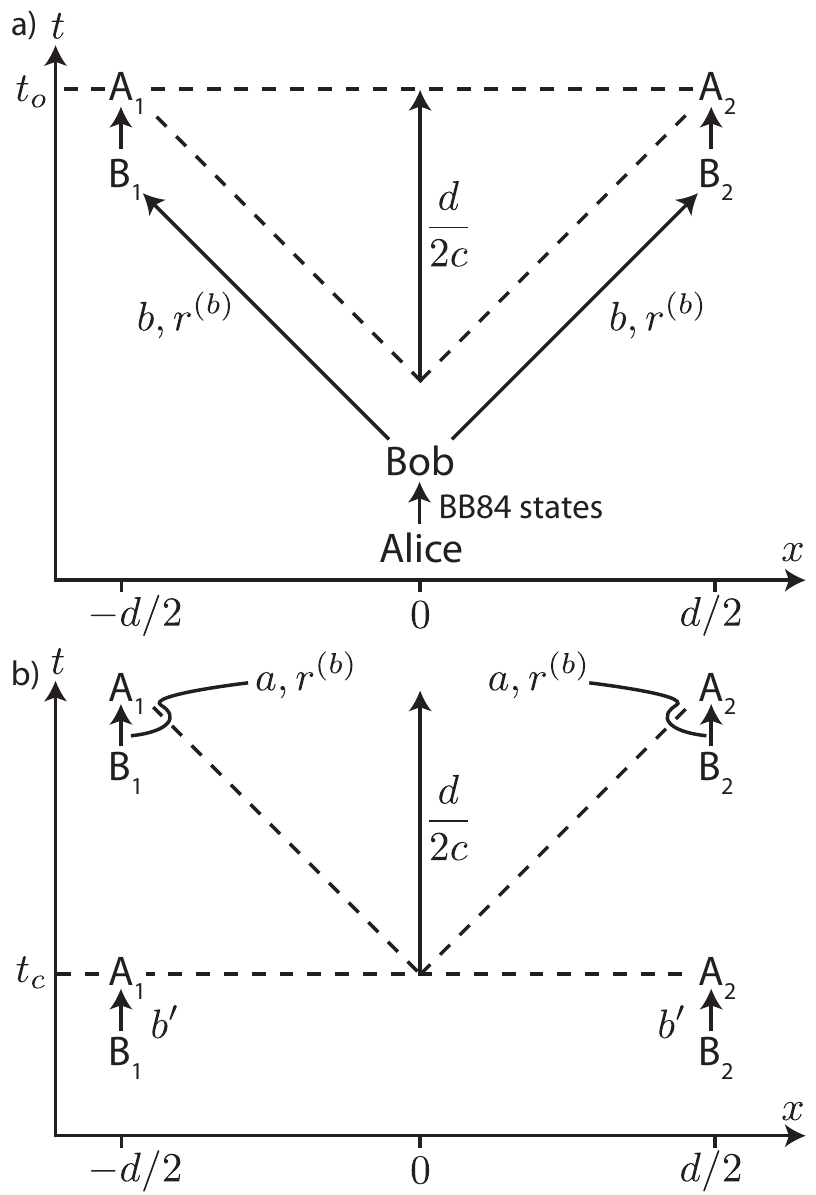}
\caption{a) Space-time ($x$-$t$) diagram of the protocol of~\cite{Kent2012a}. b) The relativistic portion of a modified version of (a). The quantum exchange between Alice and Bob happens much before and is not shown.} 
\label{protocols}
\end{figure}

Let us briefly describe the original protocol~\cite{Kent2012a}.
Fig.~\ref{protocols}a shows the evolution of the protocol in a
space-time diagram. Bob wants to commit a bit $b$ to Alice. The
protocol starts when Alice sends a group of $N$ single photons to Bob
(all photons arrive to Bob at the same time). The quantum state of
each photon is chosen at random among the four BB84 states. Bob then
immediately and simultaneously measures all photons in the $\{\ket{0},
\ket{1}\}$ basis to commit to $b=0$, or in the $\{\ket{+},\ket{-}\}$
basis to commit to $b=1$, where $\ket{\pm} =
(\ket{0}\pm\ket{1})/\sqrt{2}$. This constitutes the commitment phase.
Let us now denote ${r_i}^{(b)}$ the measurement result of the
$i^{\text{th}}$ photon. Bob then immediately sends at light-speed the
array $r^{(b)} = \{b,{r_1}^{(b)},\ldots, {r_N}^{(b)}\}$ to trusted
agents $B_1$ and $B_2$ that are symmetrically located on each side of
Bob and that are separated by a straight-line distance $d$.
Communication from Bob to each agent is encrypted using pre-shared key
material. Once agents $B_1$ and $B_2$ have received $r^{(b)}$, they
can (if they choose) later simultaneously communicate $r^{(b)}$ to the trusted agents
of Alice, $A_1$ and $A_2$, who are sitting right next to $B_1$ and
$B_2$, respectively. This opens the commitment. The commitment is
accepted when $A_1$ and $A_2$ later verify they have indeed received
the same strings at the same time, and that the measurement results
are consistent with the states she sent (hence security increases with
$N$). The duration of the commitment is $d/2c$, where $c$ is the speed
of light. Note that Bob's agents do not have to open the commitment as
soon as they receive the string $r^{(b)}$, they can wait for as long
as they wish. However, Alice has to consider that a cheating Bob could
have stored the qubits in a quantum memory and retrieved them at the
very last instant for the measurement. Hence, Alice's agents can only
be sure that Bob was committed between $t_o-d/2c$ and $t_o$, where
$t_o$ is the instant when they received~$r^{(b)}$.

This original protocol presently appears impractical for terrestrial
implementations with existing devices for direct bit commitments.
The main reason is that imperfect preparation of true
single photon states, combined with inefficient and slow
single-photon detectors, seems unlikely to allow Alice to send, and
Bob to detect, sufficiently many high-quality qubits in a
time that is negligibly small compared to $d/2c$ (i.e.~at most
21.25~ms for $d$ bounded by the Earth's diameter). 
Another problem is that it seems
difficult for an honest Bob to communicate
$r^{(b)}$ to his agents at near light speed 
unless either there is a free-space line-of-sight communication channel
between them, which requires $d \lesssim 200~$km or so on Earth,
or else neutrino pulses are sent through Earth to detectors,
which requires expensive technology not widely available. 

The protocol can be made more practical using a
delayed-choice commitment; see Fig.~\ref{protocols}-b. Specifically,
the protocol starts when Alice and Bob exchange quantum bits, but the
measurement basis of Bob, denoted $b$, is chosen at random and is
not correlated with the bit he wishes to commit to. In this way, the
quantum exchange can happen at any time and location before the
commitment phase (defined below).  In our implementation, after measuring all the
quantum bits, Bob privately communicates the results to
agents $B_1$ and $B_2$  separated by $d$,
and also tells Alice which of the qubits she sent yielded a click
in one of his detectors. This is important: Alice needs to know
before the commitment phase which qubits were detected by Bob,
otherwise there is a simple attack he can perform to break the
protocol. Bob's agents are now ready to start the commitment. For
this, in our implementation, they simultaneously send at instant $t_c$ a bit $b'=b\oplus a$
to the nearby agents $A_1$ and $A_2$ of Alice, where $a$ is the bit that Bob actually
wants to commit himself to and $b$ is the randomly committed bit. Note
that as $b$ is random, it effectively forms a one-time pad with which
Bob encrypts $a$, and hence Alice cannot learn $a$ from $b'$. At time
$t_c + d/2c$, $B_1$ and $B_2$ simultaneously reveal $a$ and $r^{(b)}$, and $A_1$
and $A_2$ can check if the measurement results are consistent with
$b$. The time $t_c$ is chosen by Bob's agents. 

Security against a malicious Alice in the ideal case is obvious, since
Bob does not reveal any information about his quantum measurement
until the opening of the commitment.  

Security against a
malicious Bob follows from special
relativity~\cite{Kent2012a,Kent2012b,Croke2012a,Kaniewski2013a}, relying
only on communication constraints that special relativity imposes.  
In relativistic cryptography, defining and analysing secure bit
commitment requires specifying the possible spacetime coordinates
of all information inputs and outputs \cite{Kent2012b}, considering
all possible spacetime locations of agents of each party.
It is thus important to note that our protocol does not guarantee to Alice that 
either $B_1$ or $B_2$, or any other agent of Bob's, knew the committed bit at time $t_c$.   
A technologically unlimited Bob could, for example locate an agent, $B_0$, halfway between $B_1$ and
$B_2$, who retains Alice's quantum states until time $t_c$, generates 
the bit $b$ at that time, and sends the measurement outcomes at light
speed to $B_1$ and $B_2$.   
$B_1$ and $B_2$ could ``blindly'' initiate the commitment at time
$t_c$ by each sending a shared random bit $b'$ 
to $A_1$ and $A_2$, and later unveil by returning the measurement
data corresponding to $b$ together with the bit $a=b' \oplus b$
effectively committed by $B_0$.   
However, our implementation {\it does} guarantee to Alice that Bob's agents 
collectively commit Bob to $a$ by time $t_c$ in the given frame, 
ensuring Bob is committed from at least $t_c$ to $t_c
+d/2c$.

It is important to realize that in the security proof it is enough to use the communication constraints imposed by special relativity. Specifically, we assume that no communication between $B_{1}$ and $B_{2}$ is possible \emph{after the commitment phase is completed}. Quantum communication gives the advantage that Alice does need to share any secret data with her agents, $A_{1}$ and $A_{2}$. On the other hand, if she tells them what quantum states were used in the commitment phase, then the dishonest Bob will be instantaneously caught cheating at the unveiling. For more information about the quantum advantage in these scenarios please refer to Appendix~\ref{quantumadvantage}.

Note that our protocol can straightforwardly be modified 
to use three (or more) separated agents of Bob's which gives certain advantages (for details refer to Appendix~\ref{three-agents}). In the two-site protocol implemented, our security proof relies on the
fact that not only is communication between $B_{1}$ and $B_{2}$ impossible
between commitment and unveiling, but no other agent of Bob's can
send classical or quantum information generated at any location after $t_c$ that
reaches them both by $t_c + (d/(2c))$.  Hence in the open phase $B_{1}$ and $B_{2}$ must
generate their answers using disjoint quantum systems, which cannot
have been jointly acted upon by any agent of Bob's after $t_c$. 

We now discuss how experimental imperfections are taken into account
in the security proof. Full details, including a step-wise description
of all protocols, are given in the supplemental material.  In practice,
qubits sent by Alice can come from attenuated laser pulses yielding a
Poisson distribution of the number of photons of mean value $\mu$.
Pulses containing two or more photons prepared in the same quantum
state can be split and measured by a malicious Bob in both bases.
Using these results only, Bob's agents are no longer bound to a single
quantum state for each declared detection, and they can choose to
reveal the result (and the basis) of their choice, provided it is the
same for $B_1$ and $B_2$, without introducing any errors in $r^{(b)}$.
Losses from Alice to Bob thus become important. Of all the pulses sent
by Alice, only a fraction will yield a click in Bob's detectors. This
fraction must therefore come from a majority of single-photon pulses,
otherwise a cheating Bob can declare a detection only when Alice's
pulse contains two or more photons at the output of her lab. Finally,
Bob can try to cheat by measuring each photon in a basis that will
minimize his probability of creating an error in $r^{(b)}$ when he
opens a bit which he was not committed to. While this probability
can be minimized, it will always be strictly greater than zero.

Let us assume that Alice sends $N$ pulses of light each containing on
average $\mu$ photons. Bob later declares that $n$ of these pulses
generated clicks in one of his detectors (and their labels, so that
Alice knows the state she sent for each detection). Let
$p_{\textrm{det}} = n/N$ be the estimated detection probability
declared by Bob. After Bob reveals $r^{(b)}$, Alice calculates the
number of bits $n'$ for which the preparation and measurement bases
were the same, as well as the number of errors $n_{\textrm{err}}$ on
these bits (with respect to the states she sent). The qubit error rate
($\text{QBER}$) is defined as $n_{\textrm{err}}/n'$. In the limit
$N\rightarrow \infty$, the protocol can be shown to be secure if
\begin{equation}
p_{\textrm{det}} > \frac{1-\e^{-\mu}(1+\mu)}{1-\frac{\text{QBER}}{\lambda}}, \label{eqn:security}
\end{equation}
where $\lambda = \frac{1}{2}\left(1-\frac{1}{\sqrt{2}} \right)\approx 0.146$. To
take into account the finite statistics, Alice and Bob both agree
before the protocol on a maximal $\text{QBER}$ value (denoted
$\delta$) and a minimal detection probability (denoted $\gamma$): only
those rounds of $N$ pulses where $p_{\textrm{det}}>\gamma$ will be
used to do a commitment. Moreover, at the end of the protocol, Alice
accepts Bob's commitment only if the $\text{QBER}<\delta$ and if the
timings of the two opening phases guarantee that no agent of Bob's
can have communicated with both $B_1$ and $B_2$ between the
commitment and unveiling.

Note that, on the one hand, suitable thresholds should be chosen to
yield a robust protocol, i.e.~the probability for it to fail when both
parties are honest should be negligibly small. On the other hand, this
increases the probability $\epsilon$ for Bob to succeed in opening any
value that he may desire after the commitment phase. The probability
$\epsilon$ can be calculated as shown in the supplemental material.

\begin{figure}[!t]
\includegraphics[width=8cm]{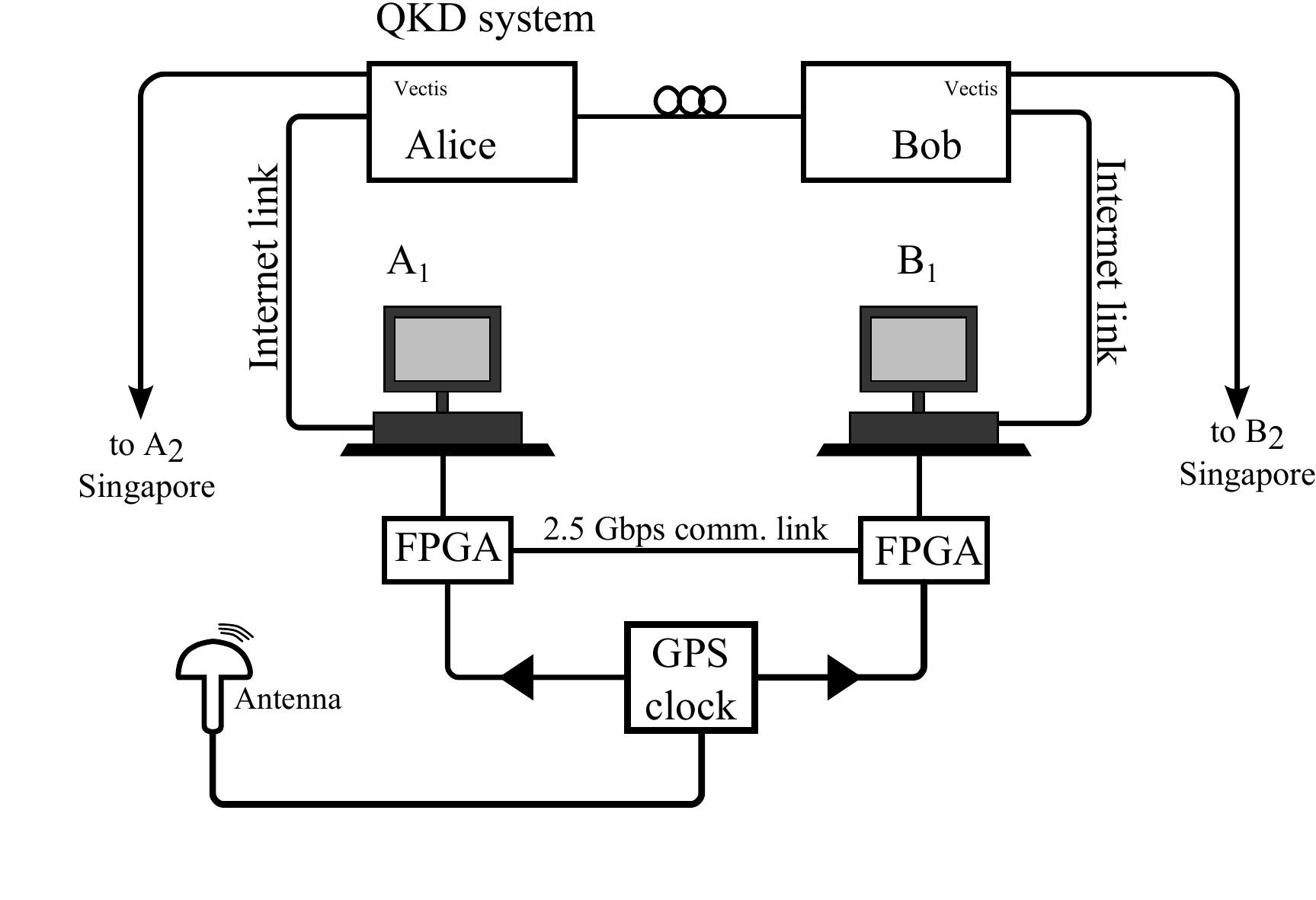}
\caption{Experimental setup located in Geneva. The setup located in Singapore is the same except for $A_1$ and $B_1$ that are replaced by $A_2$ and $B_2$, and that there is no QKD system.} 
\label{commitment_node}
\end{figure}

We implemented the modified protocol described above
(Fig.~\ref{protocols}-b). Thanks to its simplification, the quantum
part can be implemented with a quantum key distribution (QKD) system
using weak coherent pulses. In particular, we used a commercial
quantum key distribution system ``Vectis 5100'' by ID~Quantique. This
system is based on the two-way ``Plug-and-Play''
configuration~\cite{Ribordy1998a}: light pulses travel back-and-forth
from Bob to Alice through a short optical fibre with negligible loss.
When a pulse arrives on Alice side, she prepares the qubit and
attenuates the light power down to the desired $\mu$. In order to
prevent trojan-horse attacks, the incoming power is continuously
monitored~\cite{Gisin2006}. The system was installed in a office at
the University of Geneva (see Fig.~\ref{commitment_node}). The optical
setup is divided in two quantum boxes that are respectively controlled
by Alice and Bob, and that are connected by an optical fibre only. The
quantum box of Alice (or Bob) records the relevant information for the
protocol and communicates with agents $A_1$ and $A_2$ (or $B_1$ and
$B_2$) through the internet.

One important modification to how the system operates in QKD needs to
be done. Indeed, in QKD, Alice and Bob cooperate against an external
eavesdropper, while in bit commitment they are potential adversaries.
Hence, the software was modified to eliminate all the classical
communication used for key sifting except for one thing, namely the
reporting of which of the pulses sent by Alice generated a click in
one of Bob's detectors. One side-channel attack then needs to be
considered. Indeed, the probability $p_{\textrm{det}}$ of Bob's honest
apparatus might depend on the basis in which he measures. Alice
could then exploit this difference to gain information about Bob's
measurement basis, which is an issue for two-party
protocols~\cite{nelly12a}. To eliminate this attack, Bob can test his
system for an imbalance, and correct it. In practice, we monitored
during 40 hours the detection probability for each measurement basis,
and found a ratio $R = (95\pm 1)\%$. To eliminate this bias, we
programmed Bob's quantum box to declare a detection in the most
probable basis with a probability $R$.  The measurement results of Bob
are communicated to Bob's agents through the internet over a
one-time-pad encrypted channel. Similarly, the list of states sent by
Alice and for which Bob declared a detection is one-time-pad encrypted
and sent to Alice's agents.

Classical agents $A_1$ and $B_1$ were also located in an office at the
University of Geneva, while $A_2$ and $B_2$ were in an office on the
campus of the National University of Singapore (see
Fig.~\ref{commitment_node}). The straight-line distance (through the
Earth) between the two locations is about 9354~km, corresponding to a
commitment time of $d/2c = 15.6$~ms. This is close to the theoretical
maximal value of $\approx 21.2$~ms achievable with antipodal points on
the surface of the Earth. A representation of the experimental setup
is depicted in Fig.~\ref{commitment_node}. Each of the classical
agents is a standalone computer equipped with a field-programmable
gate array (FPGA) programmed to execute the necessary steps of the
protocol. Each FPGA is synchronized to universal time using a global
positioning system (GPS) clock from Oscilloquartz. Communication
between $B_1$ and $A_1$ (and similarly for $B_2$ and $A_2$) is done
over a 2.5~Gbps optical link. The time required to communicate 7000
bits (which is a typical length for the string $r^{(b)}$) is about 3
$\mu$s, which is effectively negligible compared to $15$~ms. The
opening phase is thus considered to happen instantaneously. The time at which each step of the protocol occurred was measured with a time-to-digital converter and was found to be precise to within a fraction of a microsecond.

\begin{figure}[!t]
\includegraphics[width=7cm]{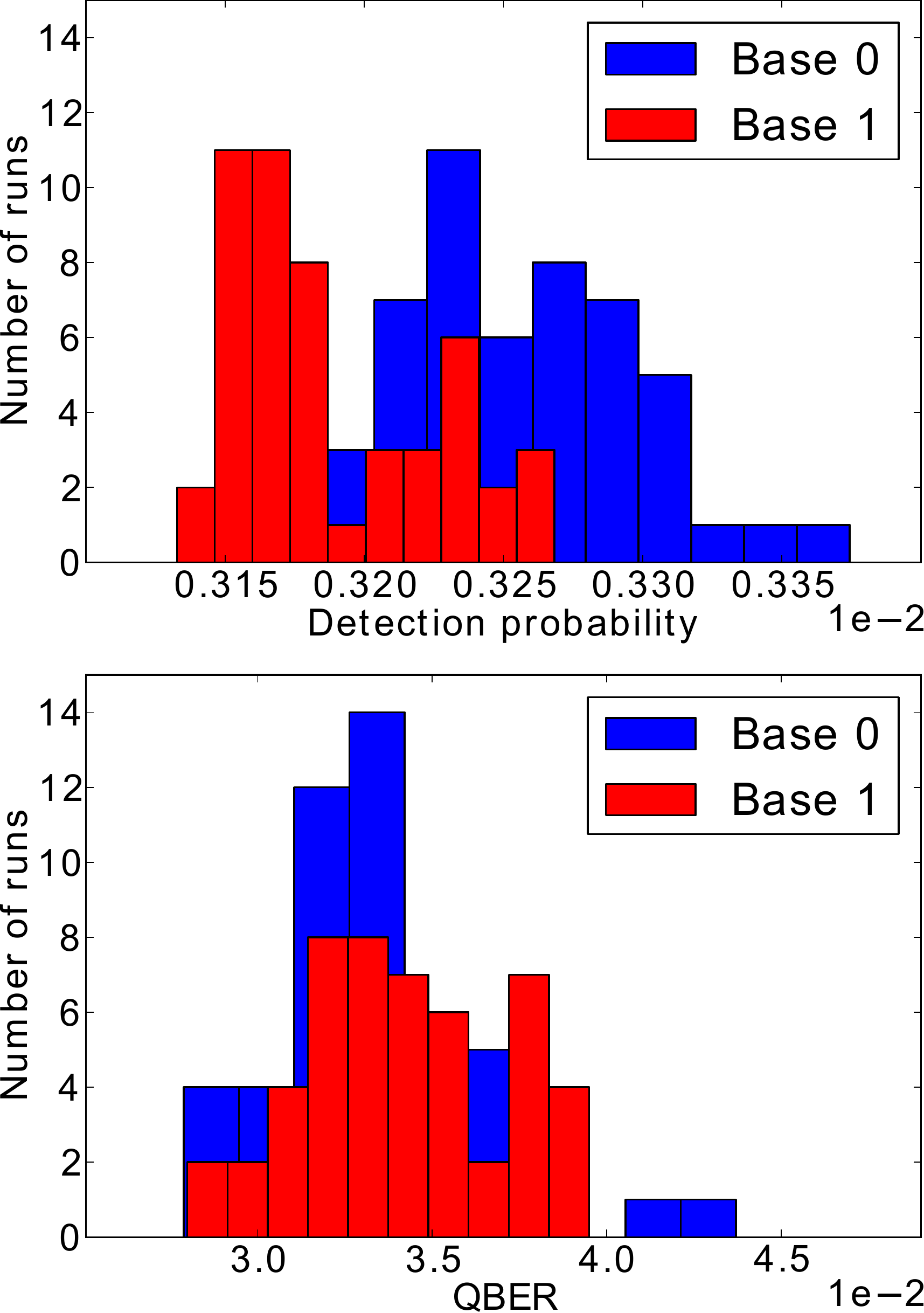}
\caption{a) Histogram of $p_{\textrm{det}}$ for one hundred commitments. The blue (red) histogram shows the commitments made in basis $b=0$ ($b=1$). b) Histogram of the observed $\text{QBERs}$ in each basis.} 
\label{Pdet_and_qber}
\end{figure}

\begin{figure}[!t]
\includegraphics[width=7cm]{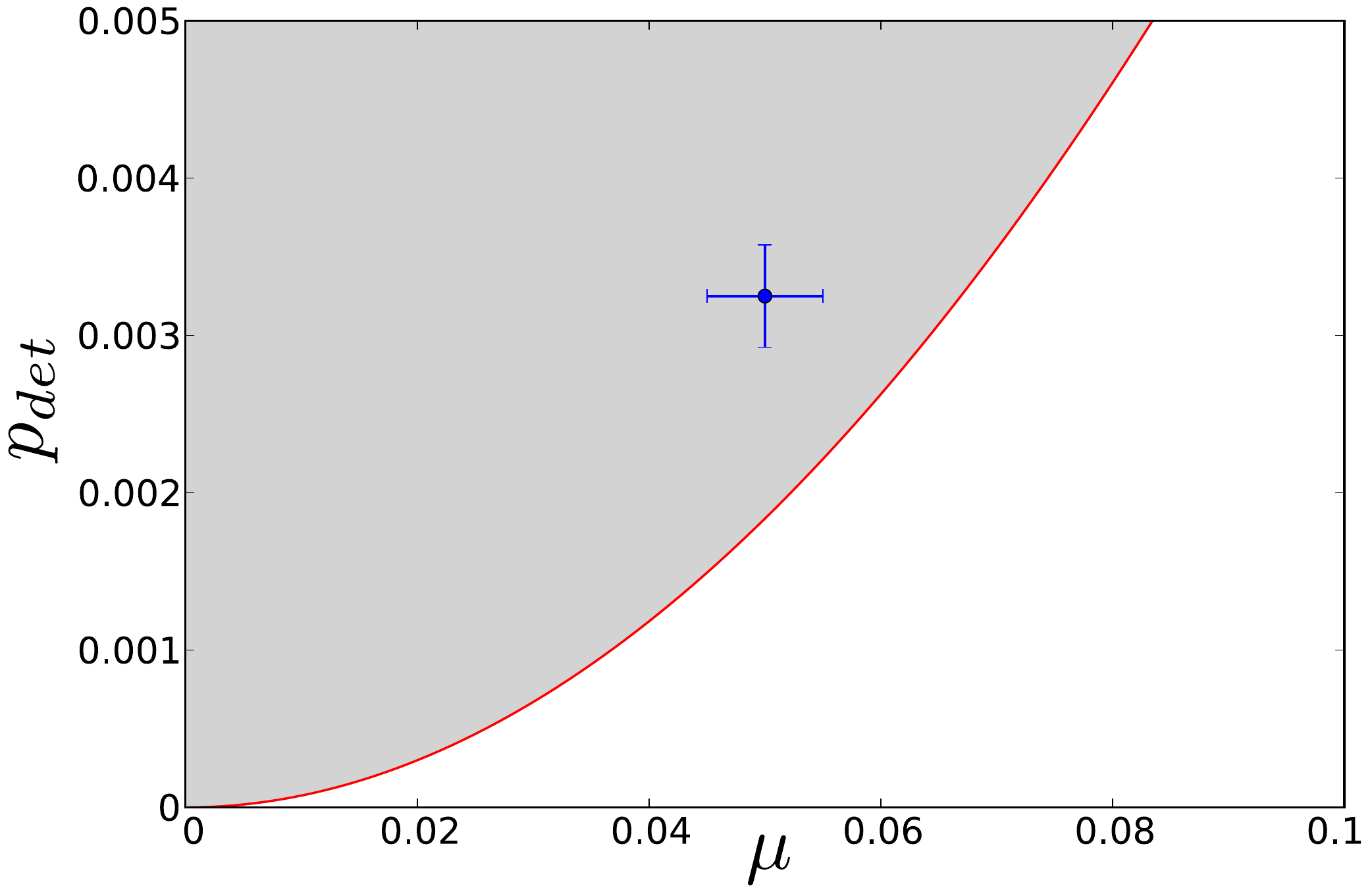}
\caption{Secure (grey) and insecure (white) regions corresponding to our experimental setup. The point corresponds to one of the commitments. The horizontal error bar correspond to the uncertainty on $\mu$, and the vertical one to the statistical uncertainty in estimating $p_{\textrm{det}}$ for this particular commitment. Note that the point is well within the secure region, which was also the case for all the other commitments performed (not shown).} 
\label{security-analysis}
\end{figure}

We realized fifty bit commitments by measuring in basis $b=0$, and fifty more in basis $b=1$, all of these with a $\mu = (5.0\pm 0.5)\times 10^{-2}$, where the uncertainty corresponds to slightly overestimated daily fluctuations. 
Figures~\ref{Pdet_and_qber}-a~and~\ref{Pdet_and_qber}-b display, respectively, the histogram of the observed $p_{\textrm{det}}$ and of QBER for these commitments. 
The optical transmission from Alice to Bob, including detector efficiency was of the order of 6\%, yielding a mean detection probability, $p_{\textrm{det}}$, of 0.32\% (this includes the contribution of dark counts and multiphoton pulses). The $\text{QBER}$ varied between 2.8 and 4.3\% and averaged at 3.4\%.
The difference between the observed detection probabilities in bases ``0'' and ``1'' are attributed to statistical fluctuations. 

Given the characterization described above, we choose to set the maximal tolerable value of the QBER to $\delta = 5\%$ which, when combined with the average $\mu$, yield a value of 0.16\% for the right-hand side of the asymptotic security condition (Eq.~\ref{eqn:security}). For the minimum tolerable detection probability $\gamma$, we choose a value of $0.2$\% for with $N=220\times10^4$ pulses sent by Alice. The number of detections declared by Bob was about 7000 per commitment. The security parameter with these numbers is $\epsilon \le 5.5\times 10^{-8}$. Figure~\ref{security-analysis} shows the asymptotic values of $p_{\textrm{det}}$ as a function of $\mu$ delimiting the secure region (in grey) from the insecure one (in white). The experimental point shown corresponds to one of the hundred commitments we realized. All the other ones are also within the secure region, which highlights the robustness of the implementation. 

We have demonstrated for the first time the possibility of
implementing practical and secure bit commitment using quantum
communication and special relativity, using a protocol with
significant quantum advantages.  Our implementation also
demonstrates the possibility of implementing such commitments 
in real time for data acquired at a single location.  This kind of system could
potentially be useful in the high-speed trading stock market where
short term commitments are sufficient. One of its main advantage is
that the quantum part of the protocol can happen at any time before
the committing phase. Hence, quantum data can be accumulated and
communicated to agents located far away. 
The commitment time can in principle be multiplied \cite{Kent1999} if Bob
commits himself to all the bits of $r^{(b)}$ instead of opening the
commitment, and this strategy can be refined \cite{Kent2005} to sustain commitments
indefinitely with constant communication rates (although security
of indefinite commitments against general quantum attacks has not been proven).
It would be an interesting theoretical and experimental
challenge to extend the unconditionally secure
commitment time significantly beyond the 15~ms obtained
here.

\paragraph*{Acknowledgments.}
This work was supported in part by the NCCR-QSIT, the European IP-SIQS, the Ministry of Education (MOE) and National
Research Foundation Singapore through a Tier 3 Grant
MOE2012-T3-1-009, and by the John Templeton Foundation.
We thank IDQuantique and Oscilloquartz for useful discussions and technical help. 
We thank N.~Walenta, C.~Ci Wen Lim, P.~Trinkler, M.~Legr\'e and D. Caselunghe for useful discussions.

\bibliographystyle{apsrev4-1}
\bibliography{bitcommitGAP,library}

\begin{thebibliography}{27}%
\makeatletter
\providecommand \@ifxundefined [1]{%
 \@ifx{#1\undefined}
}%
\providecommand \@ifnum [1]{%
 \ifnum #1\expandafter \@firstoftwo
 \else \expandafter \@secondoftwo
 \fi
}%
\providecommand \@ifx [1]{%
 \ifx #1\expandafter \@firstoftwo
 \else \expandafter \@secondoftwo
 \fi
}%
\providecommand \natexlab [1]{#1}%
\providecommand \enquote  [1]{``#1''}%
\providecommand \bibnamefont  [1]{#1}%
\providecommand \bibfnamefont [1]{#1}%
\providecommand \citenamefont [1]{#1}%
\providecommand \href@noop [0]{\@secondoftwo}%
\providecommand \href [0]{\begingroup \@sanitize@url \@href}%
\providecommand \@href[1]{\@@startlink{#1}\@@href}%
\providecommand \@@href[1]{\endgroup#1\@@endlink}%
\providecommand \@sanitize@url [0]{\catcode `\\12\catcode `\$12\catcode
  `\&12\catcode `\#12\catcode `\^12\catcode `\_12\catcode `\%12\relax}%
\providecommand \@@startlink[1]{}%
\providecommand \@@endlink[0]{}%
\providecommand \url  [0]{\begingroup\@sanitize@url \@url }%
\providecommand \@url [1]{\endgroup\@href {#1}{\urlprefix }}%
\providecommand \urlprefix  [0]{URL }%
\providecommand \Eprint [0]{\href }%
\providecommand \doibase [0]{http://dx.doi.org/}%
\providecommand \selectlanguage [0]{\@gobble}%
\providecommand \bibinfo  [0]{\@secondoftwo}%
\providecommand \bibfield  [0]{\@secondoftwo}%
\providecommand \translation [1]{[#1]}%
\providecommand \BibitemOpen [0]{}%
\providecommand \bibitemStop [0]{}%
\providecommand \bibitemNoStop [0]{.\EOS\space}%
\providecommand \EOS [0]{\spacefactor3000\relax}%
\providecommand \BibitemShut  [1]{\csname bibitem#1\endcsname}%
\let\auto@bib@innerbib\@empty
\bibitem [{\citenamefont {Blum}(1981)}]{Blum1981a}%
  \BibitemOpen
  \bibfield  {author} {\bibinfo {author} {\bibfnamefont {M.}~\bibnamefont
  {Blum}},\ }in\ \href@noop {} {\emph {\bibinfo {booktitle} {Advances in
  Cryptology: A Report on CRYPTO'81, Santa Barbara, California, USA}}}\
  (\bibinfo {year} {1981})\ pp.\ \bibinfo {pages} {11--15}\BibitemShut
  {NoStop}%
\bibitem [{\citenamefont {Broadbent}\ and\ \citenamefont
  {Tapp}(2008)}]{Broadbent2008a}%
  \BibitemOpen
  \bibfield  {author} {\bibinfo {author} {\bibfnamefont {A.}~\bibnamefont
  {Broadbent}}\ and\ \bibinfo {author} {\bibfnamefont {A.}~\bibnamefont
  {Tapp}},\ }in\ \href@noop {} {\emph {\bibinfo {booktitle} {Proceedings of the
  IAVoSS Workshop On Trustworthy Elections (WOTE 2008)}}}\ (\bibinfo {address}
  {http://eprint.iacr.org/2008/266},\ \bibinfo {year} {2008})\BibitemShut
  {NoStop}%
\bibitem [{\citenamefont {Mayers}(1997)}]{Mayers1997a}%
  \BibitemOpen
  \bibfield  {author} {\bibinfo {author} {\bibfnamefont {D.}~\bibnamefont
  {Mayers}},\ }\href {\doibase 10.1103/PhysRevLett.78.3414} {\bibfield
  {journal} {\bibinfo  {journal} {Phys. Rev. Lett.}\ }\textbf {\bibinfo
  {volume} {78}},\ \bibinfo {pages} {3414} (\bibinfo {year}
  {1997})}\BibitemShut {NoStop}%
\bibitem [{\citenamefont {Lo}\ and\ \citenamefont {Chau}(1997)}]{Lo1997a}%
  \BibitemOpen
  \bibfield  {author} {\bibinfo {author} {\bibfnamefont {H.-K.}\ \bibnamefont
  {Lo}}\ and\ \bibinfo {author} {\bibfnamefont {H.~F.}\ \bibnamefont {Chau}},\
  }\href {\doibase 10.1103/PhysRevLett.78.3410} {\bibfield  {journal} {\bibinfo
   {journal} {Phys. Rev. Lett.}\ }\textbf {\bibinfo {volume} {78}},\ \bibinfo
  {pages} {3410} (\bibinfo {year} {1997})}\BibitemShut {NoStop}%
\bibitem [{\citenamefont {D'Ariano}\ \emph {et~al.}(2007)\citenamefont
  {D'Ariano}, \citenamefont {Kretschmann}, \citenamefont {Schlingemann},\ and\
  \citenamefont {Werner}}]{Dariano2007a}%
  \BibitemOpen
  \bibfield  {author} {\bibinfo {author} {\bibfnamefont {G.~M.}\ \bibnamefont
  {D'Ariano}}, \bibinfo {author} {\bibfnamefont {D.}~\bibnamefont
  {Kretschmann}}, \bibinfo {author} {\bibfnamefont {D.}~\bibnamefont
  {Schlingemann}}, \ and\ \bibinfo {author} {\bibfnamefont {R.~F.}\
  \bibnamefont {Werner}},\ }\href {\doibase 10.1103/PhysRevA.76.032328}
  {\bibfield  {journal} {\bibinfo  {journal} {Phys. Rev. A}\ }\textbf {\bibinfo
  {volume} {76}},\ \bibinfo {pages} {032328} (\bibinfo {year}
  {2007})}\BibitemShut {NoStop}%
\bibitem [{\citenamefont {Winkler}\ \emph {et~al.}(2011)\citenamefont
  {Winkler}, \citenamefont {Tomamichel}, \citenamefont {Hengl},\ and\
  \citenamefont {Renner}}]{winkler11}%
  \BibitemOpen
  \bibfield  {author} {\bibinfo {author} {\bibfnamefont {S.}~\bibnamefont
  {Winkler}}, \bibinfo {author} {\bibfnamefont {M.}~\bibnamefont {Tomamichel}},
  \bibinfo {author} {\bibfnamefont {S.}~\bibnamefont {Hengl}}, \ and\ \bibinfo
  {author} {\bibfnamefont {R.}~\bibnamefont {Renner}},\ }\href {\doibase
  10.1103/PhysRevLett.107.090502} {\bibfield  {journal} {\bibinfo  {journal}
  {Phys. Rev. Lett.}\ }\textbf {\bibinfo {volume} {107}} (\bibinfo {year}
  {2011}),\ 10.1103/PhysRevLett.107.090502}\BibitemShut {NoStop}%
\bibitem [{\citenamefont {Aharonov}\ \emph {et~al.}(2000)\citenamefont
  {Aharonov}, \citenamefont {Ta-Shma}, \citenamefont {Vazirani},\ and\
  \citenamefont {Yao}}]{Aharonov2000a}%
  \BibitemOpen
  \bibfield  {author} {\bibinfo {author} {\bibfnamefont {D.}~\bibnamefont
  {Aharonov}}, \bibinfo {author} {\bibfnamefont {A.}~\bibnamefont {Ta-Shma}},
  \bibinfo {author} {\bibfnamefont {U.}~\bibnamefont {Vazirani}}, \ and\
  \bibinfo {author} {\bibfnamefont {A.~C.-C.}\ \bibnamefont {Yao}},\ }in\
  \href@noop {} {\emph {\bibinfo {booktitle} {Proc. of the 32nd Annual ACM
  Symp. of Theory of Computing, Portland, Oregon, USA}}}\ (\bibinfo {year}
  {2000})\ pp.\ \bibinfo {pages} {705--714}\BibitemShut {NoStop}%
\bibitem [{\citenamefont {Spekkens}\ and\ \citenamefont
  {Rudolph}(2001)}]{Spekkens2001a}%
  \BibitemOpen
  \bibfield  {author} {\bibinfo {author} {\bibfnamefont {R.~W.}\ \bibnamefont
  {Spekkens}}\ and\ \bibinfo {author} {\bibfnamefont {T.}~\bibnamefont
  {Rudolph}},\ }\href {\doibase 10.1103/PhysRevA.65.012310} {\bibfield
  {journal} {\bibinfo  {journal} {Phys. Rev. A}\ }\textbf {\bibinfo {volume}
  {65}},\ \bibinfo {pages} {012310} (\bibinfo {year} {2001})}\BibitemShut
  {NoStop}%
\bibitem [{\citenamefont {Kitaev}()}]{Kitaev2003a}%
  \BibitemOpen
  \bibfield  {author} {\bibinfo {author} {\bibfnamefont {A.}~\bibnamefont
  {Kitaev}},\ }\href@noop {} {\enquote {\bibinfo {title} {Quantum
  coin-flipping},}\ }\bibinfo {note} {Lecture delivered at the 2003 Annual
  Quantum Information Processing (QIP) Workshop, Mathematical Sciences Research
  Institute, Berkeley, CA.}\BibitemShut {Stop}%
\bibitem [{\citenamefont {Chailloux}\ and\ \citenamefont
  {Kerenidis}(2011)}]{Chailloux2011a}%
  \BibitemOpen
  \bibfield  {author} {\bibinfo {author} {\bibfnamefont {A.}~\bibnamefont
  {Chailloux}}\ and\ \bibinfo {author} {\bibfnamefont {I.}~\bibnamefont
  {Kerenidis}},\ }in\ \href@noop {} {\emph {\bibinfo {booktitle} {Foundations
  of Computer Science (FOCS), 2011 IEEE 52nd Annual Symposium on}}}\ (\bibinfo
  {year} {2011})\ pp.\ \bibinfo {pages} {354--362}\BibitemShut {NoStop}%
\bibitem [{\citenamefont {Nguyen}\ \emph {et~al.}(2008)\citenamefont {Nguyen},
  \citenamefont {Frison}, \citenamefont {Huy},\ and\ \citenamefont
  {Massar}}]{Nguyen2008a}%
  \BibitemOpen
  \bibfield  {author} {\bibinfo {author} {\bibfnamefont {A.~T.}\ \bibnamefont
  {Nguyen}}, \bibinfo {author} {\bibfnamefont {J.}~\bibnamefont {Frison}},
  \bibinfo {author} {\bibfnamefont {K.~P.}\ \bibnamefont {Huy}}, \ and\
  \bibinfo {author} {\bibfnamefont {S.}~\bibnamefont {Massar}},\ }\href
  {http://stacks.iop.org/1367-2630/10/i=8/a=083037} {\bibfield  {journal}
  {\bibinfo  {journal} {New Journal of Physics}\ }\textbf {\bibinfo {volume}
  {10}},\ \bibinfo {pages} {083037} (\bibinfo {year} {2008})}\BibitemShut
  {NoStop}%
\bibitem [{\citenamefont {Berl\`in}\ \emph {et~al.}(2011)\citenamefont
  {Berl\`in}, \citenamefont {Brassard}, \citenamefont {Bussi\`eres},
  \citenamefont {Godbout}, \citenamefont {Slater},\ and\ \citenamefont
  {Tittel}}]{Berlin2011a}%
  \BibitemOpen
  \bibfield  {author} {\bibinfo {author} {\bibfnamefont {G.}~\bibnamefont
  {Berl\`in}}, \bibinfo {author} {\bibfnamefont {G.}~\bibnamefont {Brassard}},
  \bibinfo {author} {\bibfnamefont {F.}~\bibnamefont {Bussi\`eres}}, \bibinfo
  {author} {\bibfnamefont {N.}~\bibnamefont {Godbout}}, \bibinfo {author}
  {\bibfnamefont {J.~A.}\ \bibnamefont {Slater}}, \ and\ \bibinfo {author}
  {\bibfnamefont {W.}~\bibnamefont {Tittel}},\ }\href
  {http://dx.doi.org/10.1038/ncomms1572} {\bibfield  {journal} {\bibinfo
  {journal} {Nat Commun}\ }\textbf {\bibinfo {volume} {2}},\ \bibinfo {pages}
  {561} (\bibinfo {year} {2011})}\BibitemShut {NoStop}%
\bibitem [{\citenamefont {Pappa}\ \emph {et~al.}(2013)\citenamefont {Pappa},
  \citenamefont {Jouguet}, \citenamefont {Lawson}, \citenamefont {Chailloux},
  \citenamefont {Legr\'e}, \citenamefont {Trinkler}, \citenamefont
  {Kerenidis},\ and\ \citenamefont {Diamanti}}]{Pappa2013a}%
  \BibitemOpen
  \bibfield  {author} {\bibinfo {author} {\bibfnamefont {A.}~\bibnamefont
  {Pappa}}, \bibinfo {author} {\bibfnamefont {P.}~\bibnamefont {Jouguet}},
  \bibinfo {author} {\bibfnamefont {T.}~\bibnamefont {Lawson}}, \bibinfo
  {author} {\bibfnamefont {A.}~\bibnamefont {Chailloux}}, \bibinfo {author}
  {\bibfnamefont {M.}~\bibnamefont {Legr\'e}}, \bibinfo {author} {\bibfnamefont
  {P.}~\bibnamefont {Trinkler}}, \bibinfo {author} {\bibfnamefont
  {I.}~\bibnamefont {Kerenidis}}, \ and\ \bibinfo {author} {\bibfnamefont
  {E.}~\bibnamefont {Diamanti}},\ }\href@noop {} {\bibfield  {journal}
  {\bibinfo  {journal} {arXiv:1306.3368}\ } (\bibinfo {year}
  {2013})}\BibitemShut {NoStop}%
\bibitem [{\citenamefont {Ng}\ \emph {et~al.}(2012{\natexlab{a}})\citenamefont
  {Ng}, \citenamefont {Joshi}, \citenamefont {Chen~Ming}, \citenamefont
  {Kurtsiefer},\ and\ \citenamefont {Wehner}}]{Ng2012a}%
  \BibitemOpen
  \bibfield  {author} {\bibinfo {author} {\bibfnamefont {N.~H.~Y.}\
  \bibnamefont {Ng}}, \bibinfo {author} {\bibfnamefont {S.~K.}\ \bibnamefont
  {Joshi}}, \bibinfo {author} {\bibfnamefont {C.}~\bibnamefont {Chen~Ming}},
  \bibinfo {author} {\bibfnamefont {C.}~\bibnamefont {Kurtsiefer}}, \ and\
  \bibinfo {author} {\bibfnamefont {S.}~\bibnamefont {Wehner}},\ }\href
  {http://dx.doi.org/10.1038/ncomms2268} {\bibfield  {journal} {\bibinfo
  {journal} {Nat Commun}\ }\textbf {\bibinfo {volume} {3}},\ \bibinfo {pages}
  {1326} (\bibinfo {year} {2012}{\natexlab{a}})}\BibitemShut {NoStop}%
\bibitem [{\citenamefont {Kent}(1999)}]{Kent1999}%
  \BibitemOpen
  \bibfield  {author} {\bibinfo {author} {\bibfnamefont {A.}~\bibnamefont
  {Kent}},\ }\href {\doibase 10.1103/PhysRevLett.83.1447} {\bibfield  {journal}
  {\bibinfo  {journal} {Phys. Rev. Lett.}\ }\textbf {\bibinfo {volume} {83}},\
  \bibinfo {pages} {1447} (\bibinfo {year} {1999})}\BibitemShut {NoStop}%
\bibitem [{\citenamefont {Kent}(2005)}]{Kent2005}%
  \BibitemOpen
  \bibfield  {author} {\bibinfo {author} {\bibfnamefont {A.}~\bibnamefont
  {Kent}},\ }\href@noop {} {\bibfield  {journal} {\bibinfo  {journal} {Journal
  of Cryptology}\ }\textbf {\bibinfo {volume} {18}},\ \bibinfo {pages} {313}
  (\bibinfo {year} {2005})}\BibitemShut {NoStop}%
\bibitem [{\citenamefont {Kent}(2011)}]{Kent2011}%
  \BibitemOpen
  \bibfield  {author} {\bibinfo {author} {\bibfnamefont {A.}~\bibnamefont
  {Kent}},\ }\href@noop {} {\bibfield  {journal} {\bibinfo  {journal} {New
  Journal of Physics}\ }\textbf {\bibinfo {volume} {13}},\ \bibinfo {pages}
  {113015} (\bibinfo {year} {2011})}\BibitemShut {NoStop}%
\bibitem [{\citenamefont {Kent}(2012{\natexlab{a}})}]{Kent2012a}%
  \BibitemOpen
  \bibfield  {author} {\bibinfo {author} {\bibfnamefont {A.}~\bibnamefont
  {Kent}},\ }\href {\doibase 10.1103/PhysRevLett.109.130501} {\bibfield
  {journal} {\bibinfo  {journal} {Phys. Rev. Lett.}\ }\textbf {\bibinfo
  {volume} {109}},\ \bibinfo {pages} {130501} (\bibinfo {year}
  {2012}{\natexlab{a}})}\BibitemShut {NoStop}%
\bibitem [{\citenamefont {Croke}\ and\ \citenamefont
  {Kent}(2012)}]{Croke2012a}%
  \BibitemOpen
  \bibfield  {author} {\bibinfo {author} {\bibfnamefont {S.}~\bibnamefont
  {Croke}}\ and\ \bibinfo {author} {\bibfnamefont {A.}~\bibnamefont {Kent}},\
  }\href {\doibase 10.1103/PhysRevA.86.052309} {\bibfield  {journal} {\bibinfo
  {journal} {Phys. Rev. A}\ }\textbf {\bibinfo {volume} {86}},\ \bibinfo
  {pages} {052309} (\bibinfo {year} {2012})}\BibitemShut {NoStop}%
\bibitem [{\citenamefont {Kaniewski}\ \emph {et~al.}(2013)\citenamefont
  {Kaniewski}, \citenamefont {Tomamichel}, \citenamefont {Hanggi},\ and\
  \citenamefont {Wehner}}]{Kaniewski2013a}%
  \BibitemOpen
  \bibfield  {author} {\bibinfo {author} {\bibfnamefont {J.}~\bibnamefont
  {Kaniewski}}, \bibinfo {author} {\bibfnamefont {M.}~\bibnamefont
  {Tomamichel}}, \bibinfo {author} {\bibfnamefont {E.}~\bibnamefont {Hanggi}},
  \ and\ \bibinfo {author} {\bibfnamefont {S.}~\bibnamefont {Wehner}},\ }\href
  {\doibase 10.1109/TIT.2013.2247463} {\bibfield  {journal} {\bibinfo
  {journal} {Information Theory, IEEE Transactions on}\ }\textbf {\bibinfo
  {volume} {59}},\ \bibinfo {pages} {4687} (\bibinfo {year}
  {2013})}\BibitemShut {NoStop}%
\bibitem [{\citenamefont {Kent}(2012{\natexlab{b}})}]{Kent2012b}%
  \BibitemOpen
  \bibfield  {author} {\bibinfo {author} {\bibfnamefont {A.}~\bibnamefont
  {Kent}},\ }\href@noop {} {\bibfield  {journal} {\bibinfo  {journal}
  {Classical and Quantum Gravity}\ }\textbf {\bibinfo {volume} {29}},\ \bibinfo
  {pages} {224013} (\bibinfo {year} {2012}{\natexlab{b}})}\BibitemShut
  {NoStop}%
\bibitem [{\citenamefont {Ribordy}\ \emph {et~al.}(1998)\citenamefont
  {Ribordy}, \citenamefont {Gautier}, \citenamefont {Gisin}, \citenamefont
  {Guinnard},\ and\ \citenamefont {Zbinden}}]{Ribordy1998a}%
  \BibitemOpen
  \bibfield  {author} {\bibinfo {author} {\bibfnamefont {G.}~\bibnamefont
  {Ribordy}}, \bibinfo {author} {\bibfnamefont {J.~D.}\ \bibnamefont
  {Gautier}}, \bibinfo {author} {\bibfnamefont {N.}~\bibnamefont {Gisin}},
  \bibinfo {author} {\bibfnamefont {O.}~\bibnamefont {Guinnard}}, \ and\
  \bibinfo {author} {\bibfnamefont {H.}~\bibnamefont {Zbinden}},\ }\href@noop
  {} {\bibfield  {journal} {\bibinfo  {journal} {Electronics Letters}\ }\textbf
  {\bibinfo {volume} {34}},\ \bibinfo {pages} {2116} (\bibinfo {year}
  {1998})}\BibitemShut {NoStop}%
\bibitem [{\citenamefont {Gisin}\ \emph {et~al.}(2006)\citenamefont {Gisin},
  \citenamefont {Fasel}, \citenamefont {Kraus}, \citenamefont {Zbinden},\ and\
  \citenamefont {Ribordy}}]{Gisin2006}%
  \BibitemOpen
  \bibfield  {author} {\bibinfo {author} {\bibfnamefont {N.}~\bibnamefont
  {Gisin}}, \bibinfo {author} {\bibfnamefont {S.}~\bibnamefont {Fasel}},
  \bibinfo {author} {\bibfnamefont {B.}~\bibnamefont {Kraus}}, \bibinfo
  {author} {\bibfnamefont {H.}~\bibnamefont {Zbinden}}, \ and\ \bibinfo
  {author} {\bibfnamefont {G.}~\bibnamefont {Ribordy}},\ }\href {\doibase
  10.1103/PhysRevA.73.022320} {\bibfield  {journal} {\bibinfo  {journal}
  {Physical Review A}\ }\textbf {\bibinfo {volume} {73}},\ \bibinfo {pages}
  {022320} (\bibinfo {year} {2006})}\BibitemShut {NoStop}%
\bibitem [{\citenamefont {Ng}\ \emph {et~al.}(2012{\natexlab{b}})\citenamefont
  {Ng}, \citenamefont {Joshi}, \citenamefont {Ming}, \citenamefont
  {Kurtsiefer},\ and\ \citenamefont {Wehner}}]{nelly12a}%
  \BibitemOpen
  \bibfield  {author} {\bibinfo {author} {\bibfnamefont {N.~H.~Y.}\
  \bibnamefont {Ng}}, \bibinfo {author} {\bibfnamefont {S.~K.}\ \bibnamefont
  {Joshi}}, \bibinfo {author} {\bibfnamefont {C.~C.}\ \bibnamefont {Ming}},
  \bibinfo {author} {\bibfnamefont {C.}~\bibnamefont {Kurtsiefer}}, \ and\
  \bibinfo {author} {\bibfnamefont {S.}~\bibnamefont {Wehner}},\ }\href
  {\doibase 10.1038/ncomms2268} {\bibfield  {journal} {\bibinfo  {journal}
  {Nat. Commun.}\ }\textbf {\bibinfo {volume} {3}},\ \bibinfo {pages} {1326}
  (\bibinfo {year} {2012}{\natexlab{b}})}\BibitemShut {NoStop}%
\bibitem [{\citenamefont {Chernoff}(1952)}]{chernoff52}%
  \BibitemOpen
  \bibfield  {author} {\bibinfo {author} {\bibfnamefont {H.}~\bibnamefont
  {Chernoff}},\ }\href@noop {} {\bibfield  {journal} {\bibinfo  {journal} {Ann.
  Math. Stat.}\ }\textbf {\bibinfo {volume} {23}},\ \bibinfo {pages} {493}
  (\bibinfo {year} {1952})}\BibitemShut {NoStop}%
\bibitem [{\citenamefont {Ben-Or}\ \emph {et~al.}(1988)\citenamefont {Ben-Or},
  \citenamefont {Goldwasser}, \citenamefont {Kilian},\ and\ \citenamefont
  {Widgerson}}]{benor:bc}%
  \BibitemOpen
  \bibfield  {author} {\bibinfo {author} {\bibfnamefont {M.}~\bibnamefont
  {Ben-Or}}, \bibinfo {author} {\bibfnamefont {S.}~\bibnamefont {Goldwasser}},
  \bibinfo {author} {\bibfnamefont {J.}~\bibnamefont {Kilian}}, \ and\ \bibinfo
  {author} {\bibfnamefont {A.}~\bibnamefont {Widgerson}},\ }in\ \href {\doibase
  10.1145/62212.62223} {\emph {\bibinfo {booktitle} {Proc. ACM STOC}}}\
  (\bibinfo  {publisher} {ACM Press},\ \bibinfo {address} {New York, New York,
  USA},\ \bibinfo {year} {1988})\ pp.\ \bibinfo {pages} {113--131}\BibitemShut
  {NoStop}%
\bibitem [{\citenamefont {Simard}(2007)}]{bcThesis}%
  \BibitemOpen
  \bibfield  {author} {\bibinfo {author} {\bibfnamefont {J.-R.}\ \bibnamefont
  {Simard}},\ }\emph {\bibinfo {title} {{Classical and Quantum Strategies for
  Bit Commitment Schemes in the Two-Prover Model}}},\ \href@noop {} {\bibinfo
  {type} {Master's thesis}},\ \bibinfo  {school} {McGill University} (\bibinfo
  {year} {2007})\BibitemShut {NoStop}%
\end{thebibliography}%

\clearpage
\onecolumngrid
\appendix

\newcommand{\ave}[1]{\langle #1 \rangle}
\renewcommand{\bra}[1]{\langle #1 \hspace{1pt} |}
\renewcommand{\ket}[1]{| \hspace{1pt} #1 \rangle}
\newcommand{\braket}[2]{\langle #1 \hspace{1pt} | \hspace{1pt} #2 \rangle}
\newcommand{\braketq}[1]{\braket{#1}{#1}}
\newcommand{\ketbra}[2]{| \hspace{1pt} #1 \rangle \langle #2 \hspace{1pt} |}
\newcommand{\ketbras}[3]{| \hspace{1pt} #1 \rangle_{#3} \langle #2 \hspace{1pt} |}
\newcommand{\ketbraq}[1]{\ketbra{#1}{#1}}
\newcommand{\bramatket}[3]{\langle #1 \hspace{1pt} | #2 | \hspace{1pt} #3 \rangle}
\newcommand{\bramatketq}[2]{\bramatket{#1}{#2}{#1}}
\newcommand{\fulmix}[1]{\frac{\mathbb{1}_{#1}}{#1}}
\newcommand{\nbox}[2][9]{\hspace{#1pt} \mbox{#2} \hspace{#1pt}}

\newcommand{\norm}[2][]{#1| \! #1| #2 #1| \! #1|}
\newcommand{\abs}[2][]{#1| #2 #1|}
%
%
\newcommand{\tran}[0]{^\textnormal{T}}

\newcommand{\amsbb}[1]{\mathbb{#1}}

\newcommand{\red}[1]{\textcolor{red}{#1}}
\newcommand{\blue}[1]{\textcolor{blue}{#1}}
\newcommand{\com}[1]{\textcolor{red}{[#1]}}
\newcommand{\pass}{^{\textnormal{pass}}}
\newcommand{\fail}{^{\textnormal{fail}}}
\newcommand{\citneed}{\textcolor{red}{[CITATION NEEDED]}}
\newcommand{\colbold}[2]{\textcolor{#1}{\textbf{#2}}}
\newcommand{\utb}[1]{\textcolor{blue}{[used to be : #1]}}
\newcommand{\comres}[2]
{
\item \textcolor{red}{#1}\\
#2
}

\newcommand{\rep}[2]{We have replaced ``#1'' with ``#2'' as suggested.}
\newcommand{\chanmade}[0]{We have made the suggested change.}

\newcommand{\statpro}[2]
{
\begin{equation*} #1 \end{equation*}
\ifthenelse{\equal{\displayproofs}{1}}{\begin{proof} #2 \end{proof}}{}
}

\def \includeprotocol {0}
\def \includetikz {0}
\def \includecitations {1}

\newcommand{\cA}{\mathcal{A}}
\newcommand{\cB}{\mathcal{B}}
\newcommand{\cC}{\mathcal{C}}
\newcommand{\cD}{\mathcal{D}}
\newcommand{\cE}{\mathcal{E}}
\newcommand{\cF}{\mathcal{F}}
\newcommand{\cG}{\mathcal{G}}
\newcommand{\cH}{\mathcal{H}}
\newcommand{\cI}{\mathcal{I}}
\newcommand{\cJ}{\mathcal{J}}
\newcommand{\cK}{\mathcal{K}}
\newcommand{\cL}{\mathcal{L}}
\newcommand{\cM}{\mathcal{M}}
\newcommand{\cN}{\mathcal{N}}
\newcommand{\cO}{\mathcal{O}}
\newcommand{\cP}{\mathcal{P}}
\newcommand{\cQ}{\mathcal{Q}}
\newcommand{\cR}{\mathcal{R}}
\newcommand{\cS}{\mathcal{S}}
\newcommand{\cT}{\mathcal{T}}
\newcommand{\cU}{\mathcal{U}}
\newcommand{\cV}{\mathcal{V}}
\newcommand{\cW}{\mathcal{W}}
\newcommand{\cX}{\mathcal{X}}
\newcommand{\cY}{\mathcal{Y}}
\newcommand{\cZ}{\mathcal{Z}}

\newcommand{\sA}{\mathscr{A}}
\newcommand{\sB}{\mathscr{B}}
\newcommand{\sC}{\mathscr{C}}
\newcommand{\sD}{\mathscr{D}}
\newcommand{\sE}{\mathscr{E}}
\newcommand{\sF}{\mathscr{F}}
\newcommand{\sG}{\mathscr{G}}
\newcommand{\sH}{\mathscr{H}}
\newcommand{\sL}{\mathscr{L}}
\newcommand{\sM}{\mathscr{M}}
\newcommand{\sN}{\mathscr{N}}
\newcommand{\sS}{\mathscr{S}}
\newcommand{\sT}{\mathscr{T}}
\newcommand{\sX}{\mathscr{X}}
\newcommand{\sY}{\mathscr{X}}
\newcommand{\sZ}{\mathscr{Z}}

\newcommand{\printprotocol}[1]
{
\begin{enumerate}

\item In every round Alice's agent Anne ($A_{1}$) generates two uniformly random bits, $x_{k}, \theta_{k} \in \{0, 1\}$, and sends a pulse of photons in the state $\ket{\psi_{x}^{\theta}}$ to Bill ($B_{1}$).

\item (commit) Bill  chooses the bit he wants to commit to, denoted by $b$. He measures all the incoming photons in the computational basis if $b = 0$ or in the Hadamard basis if $b = 1$. Let $y$ denote the string of measurement outcomes.
\ifthenelse{#1 > 0}
{
\item (backreporting) Let $\cM \subseteq [n]$ be the \emph{valid} set,
  namely the rounds in which Bill detected a click. Bill sends $\cM$
  to Anne. Anne checks whether $m = \abs{\cM} \geq \gamma n $, otherwise she aborts the protocol. Denote the classical measurement outcomes on the valid set by $y \in \{0, 1\}^{m}$.
}
{}
Bill sends $b$ and $y$ to Brian ($B_{2}$).

\item (open) To open the commitment Bill sends $b$ and $y$ to Anne and Brian does the same to Amy ($A_{2}$). Let us denote the respective variables by $b'$ and $y'$.

\item Amy forwards the string she received from Brian to Anne.

\item (verify)
\ifthenelse{#1 > 0}
{
Let $S$ and $T$ be the following sets
\begin{align*}
S = \{k \in \cM : \theta_{k} = 0\},\\
T = \{k \in \cM : \theta_{k} = 1\}.
\end{align*}

Note that these are essentially the sets defined in Section~\ref{sec:notation} but truncated to the valid rounds only.
}
{}
Let $x_{S(T)}$ be the substring of $x$ on the set $S(T)$
\ifthenelse{#1 = 0}
{
as defined in Section~\ref{sec:notation}.
}
{}
Alice accepts the unveiling if the following three conditions are satisfied:
\begin{itemize}
\item Bill and Brian unveil the same value, $b = b'$.
\item Bill and Brian submit the same outcome string, $y = y'$.
\item The string $y$ is consistent with Alice's original encoding. More specifically, if $b = 0$ we require $\dham(x_{S}, y_{S}) \leq \delta$, whereas if $b = 1$ we require $\dham(x_{T}, y_{T}) \leq \delta$.
\end{itemize}
\end{enumerate}
}

\section{Security Analysis}

\subsection{Outline}

%
We begin our security analysis by describing how we model the honest and dishonest scenarios, i.e.~what assumptions are made on the devices used to execute the protocol. In Section~\ref{sec:fault-tolerant-rbc} we describe a generalized protocol, in which the states emitted by Alice in the commit phase do not need to be \emph{exactly} the four BB84 states and Bob's answer in the open phase does not need to match \emph{exactly} the string encoded by Alice. For this fault-tolerant RBC protocol we present a security proof and compute an explicit security bound as a function of the number of qubits exchanged in the commit phase. This allows us to deal with errors on the communication channel in the experiment, but not yet with losses. Note that we do not simply extend the analysis of~\cite{Kaniewski2013a}, but in fact our analysis is based on entirely new techniques that can - as a byproduct - also be used to obtain better parameters for the idealised setting. In Section~\ref{sec:fault-tolerant-rbc-with-backreporting} we finally make our protocol robust against errors \emph{and} losses.
In particular, we show how the fault-tolerant RBC can be extended to deal with the presence of losses. We investigate how good the devices owned by the honest parties need to be to result in a protocol that is both robust and secure in the asymptotic limit. Finally, we show how to compute explicit security bounds for a finite-length protocol.

\subsection{Modelling the experiment}
\label{sec:modelling-the-experiment}

Our model hinges on the following assumptions:
\begin{enumerate}

\item Alice uses a weak-coherent source with phase randomization.

\item Every photon is subject to independent noise and loss processes.

\item The detectors used by Bob for different bases have the same efficiency (independent of the state).

\end{enumerate}

In the following we elaborate on what thes assumptions mean for the honest and dishonest scenarios.

\subsubsection{The honest scenario}

Assuming that Alice uses a weak-coherent source with phase randomisation the ensemble emitted by the source is described by

\begin{gather*}
\rho = \sum_{r = 0}^{\infty} p_{r} \ketbraq{r},\\
\nbox{where} p_{r} = e^{- \mu} \cdot \frac{\mu^{r}}{r!},
\end{gather*}
$\mu$ is the average number of photons per pulse and $\ket{r}$ is the Fock state of $r$ photons. To model noise and losses between Alice and Bob for every photon we introduce an independent, binary random variable. Let $\eta$ be the detection efficiency: every photon emitted by the source gets detected on Bob's side with probability $\eta$, otherwise it is lost. Applying such a process to the original state simply affects the probability distribution: the probability of Bob detecting $r$ photons equals
\begin{equation}
\label{eq:pr-definition}
p_{r}(\mu, \eta) = e^{-\mu \eta} \cdot \frac{(\mu \eta)^{r}}{r!}.
\end{equation}
Now, suppose that $n$ rounds are played (these are independent due to Assumption 2) and let $N_{r}$ be the random variable counting the number of rounds in which exactly $t$ photons were received. Then
\begin{equation}
\label{eq:nt-and-expectation-value}
\Pr[N_{r} = k] = {n \choose k} p_{r}^{k} (1 - p_{r})^{n - k} \nbox{and} \amsbb{E}[N_{r}] = p_{r} n.
\end{equation}

We model bit-flip errors in a similar way. For every photon detected on Bob's side and measured in the correct basis we introduce a bit flip error with probability $Q$.

\subsubsection{Dishonest scenarios}

In the case of one party being dishonest we need to assume that  his/her devices are perfect\,---\,we want to achieve security against all-powerful adversaries. On the other hand, our assumptions still apply to the honest party.

If Alice is honest then we model her source as described before but
now no losses or errors are guaranteed. Hence, what Bob sees are light
pulses containing a Poisson-distributed (with parameter $\mu$) number
of photons. Moreover, we assume that Bob can do a perfect
photon-counting experiment, which might help him to cheat. 
Bob can locate any number of agents anywhere in space-time, send 
classical or quantum information, including the quantum states
sent by Alice, securely between his agents at
light speed, and carry out arbitrary quantum operations on Alice's
states and any ancillae. 

If Bob is honest then we make no assumptions on the state emitted by
Alice but we assume that the detectors Bob is using for the two
different bases have the same efficiency (and that it does not depend
on the state prepared by Alice).

\subsection{Notation}
\label{sec:notation}

We also fix some notation.
Let $[n] = \{1, 2, \ldots, n\}$ and $\cC_{n} = \{0, 1\}^{n}$. For $\theta \in \cC_{n}$ define the following sets
\begin{align*}
S = \{k \in [n] : \theta_{k} = 0\},\\
T = \{k \in [n] : \theta_{k} = 1\}.
\end{align*}
Note that $S \cup T = [n]$ and $S \cap T = \varnothing$. The (normalised) Hamming distance between two strings $x, y \in C_{n}$ is 
\begin{equation*}
\dham(x, y) := \frac{1}{n} \abs[\big]{\{k \in [n]: x_{k} \neq y_{k}\}}.
\end{equation*}
The Hamming weight is defined as the distance from the string of all zeroes:
\begin{equation*}
\wham(x) := \dham(x, 00 \ldots 0).
\end{equation*}

\subsection{Fault-tolerant RBC}
\label{sec:fault-tolerant-rbc}
The protocol presented in this section is essentially the original protocol in~\cite{Kent2012a} with the following two extensions:
\begin{itemize}
\item In the original protocol Alice is required to create random
  BB84 states. In other words, she is required to generate two
  uniformly random bits, $x, \theta \in \{0, 1\}$, and create the
  state $H^{\theta} \ket{x}$ of a qubit. We relax this constraint and
  simply require the two states for the same value of $\theta$ to be
  orthogonal: if $\ket{\psi_{x}^{\theta}}$ is the state corresponding
  to the random variables being $x$ and $\theta$ then we require
\begin{equation*}
\braket{\psi_{0}^{0}}{\psi_{1}^{0}} = \braket{\psi_{0}^{1}}{\psi_{1}^{1}} = 0,
\end{equation*}
which for qubit states implies
\begin{equation}
\label{eq:constraint-on-states}
\ketbraq{\psi_{0}^{0}} + \ketbraq{\psi_{1}^{0}} = \ketbraq{\psi_{0}^{1}} + \ketbraq{\psi_{1}^{1}} = \mathbb{1}.
\end{equation}
The assumption that each pair forms a basis means that any prepare-and-measure scheme can be turned into an entanglement-based one. It will become evident from the security proof that the maximally incompatible states (e.g.~the BB84 states) are the optimal choice from the security point of view.

\item The strings that Bob's agents, Bill ($B_{1}$) and Brian ($B_{2}$), unveil in the open phase do not have to match Alice's string exactly. The opening will be accepted as long as Bill ($B_{1}$) and Brian ($B_{2}$) unveil the same string and it is close (in terms of Hamming distance) to the one measured by Alice.

\end{itemize}

\subsubsection{The protocol}
Let $\delta$ be the error threshold and $n$ be the number of rounds that Alice and Bob agree to play.
\printprotocol{0}

\subsubsection{Security for honest Alice}
\label{sec:security-proof}
A dishonest Bob can locate agents anywhere in space-time and transmit
classical and quantum information between them securely at light
speed.   However, the relativistic signalling constraints implied
by the timings in our protocol mean that the quantum states
controlled respectively by Bill and Brian at the unveiling stage
belong to disjoint Hilbert spaces that cannot have interacted 
since the commitment time $t_c$.   We can thus restrict attention
to the Hilbert spaces defining Bill and Brian's possible 
states at unveiling and to Bill and Brian's actions on
these spaces.    
 
We use the security model discussed in~\cite{Kaniewski2013a}, in
particular, we employ the global command scenario. 
Suppose that at the beginning of the open phase Bill and Brian receive a message asking them to unveil the bit $b$ and let $p_{b}$ be the probability that they succeed. We will show that for any strategy
\begin{equation}
\label{eq:security-definition}
p_{0} + p_{1} \leq 1 + \varepsilon,
\end{equation}
where $\varepsilon$ decays exponentially in $n$, the number of rounds played.

The most general strategy by cheating Bob is to apply a quantum
channel that ``splits up'' the quantum state into two parts: one of
which is sent to Bill at or before unveiling, while the other is sent
to Brian at or before unveiling. 
In the open phase, each of the agents will know which bit they are
trying to unveil and will perform an arbitrary measurement on their
respective systems to find out what answers they should send to
Alice's agents. 

Hence, the cheating strategy consists of a splitting channel and four
POVMs (each agent has two different measurements depending on what
they are trying to unveil). Let the measurement operators
corresponding to Bill (Brian) trying to unveil $b$ be
$\{P_{y}^{b}\}_{y \in \cC_{n}}$ ($\{Q_{y}^{b}\}_{y \in \cC_{n}}$).
Note that we can without loss of generality assume that these
measurements are projective (by Neumark's dilation theorem since we
make no assumptions on the dimension of the quantum state).

Moreover, let us purify the protocol. Generating one of the four
states satisfying~\eqref{eq:constraint-on-states} uniformly at random
is equivalent to creating an EPR pair, and measuring one half of it in
one of the two bases with equal probability. It is clear that the two
protocols are equivalent from the security point of view but the
purified version turns out to be easier to analyse.

Let $\theta \in \cC_{n}$ be the basis string that Alice picks
uniformly at random. Then her measurement operators take the form
$\ket{x^{\theta}} = \bigotimes_{k = 1}^{n}
\ket{\psi_{x_{k}}^{\theta_{k}}}$ for $x \in \cC_{n}$. Now, the checks
that Alice performs can be written as projectors. Let
$\Pi_{b}^{\theta}$ be the projector corresponding to Alice accepting
the unveiling of $b$ given that her basis string is $\theta$. Then
\begin{align*}
\Pi_{0}^{\theta} &= \sum_{x \in \cC_{n}} \ketbraq{x^{\theta}} \otimes \sum_{\substack{y \in \cC_{n}\\ \dham(x_{S}, y_{S}) \leq \delta}} P_{y}^{0} \otimes Q_{y}^{0},\\
\Pi_{1}^{\theta} &= \sum_{x \in \cC_{n}} \ketbraq{x^{\theta}} \otimes \sum_{\substack{y \in \cC_{n}\\ \dham(x_{T}, y_{T}) \leq \delta}} P_{y}^{1} \otimes Q_{y}^{1}.
\end{align*}
Let $\rho$ be the tripartite state shared between Alice, Bill and Brian. Then the probability of successfully unveiling $b$ can be written as
\begin{equation*}
p_{b} = 2^{-n} \sum_{\theta \in \cC_{n}} \tr(\rho \Pi_{b}^{\theta}).
\end{equation*}
We want to bound $p_{0} + p_{1}$, which corresponds to
\begin{equation}
\label{eq:p0-plus-p1}
p_{0} + p_{1} = 2^{-n} \sum_{\theta \in \cC_{n}} \tr \big( \rho [\Pi_{0}^{\theta} + \Pi_{1}^{\theta}] \big).
\end{equation}
Adding the two projectors up yields
\begin{equation*}
\Pi_{0}^{\theta} + \Pi_{1}^{\theta} = \sum_{x \in \cC_{n}} \ketbraq{x^{\theta}} \otimes \Bigg[ \sum_{\substack{y \in \cC_{n}\\ \dham(x_{S}, y_{S}) \leq \delta}} P_{y}^{0} \otimes Q_{y}^{0} \quad + \sum_{\substack{y \in \cC_{n}\\ \dham(x_{T}, y_{T}) \leq \delta}} P_{y}^{1} \otimes Q_{y}^{1} \Bigg].
\end{equation*}
The terms in the square bracket can be upper bounded by replacing one of the measurement operators by the identity matrix. Therefore,
\begin{gather*}
\sum_{\substack{y \in \cC_{n}\\ \dham(x_{S}, y_{S}) \leq \delta}} P_{y}^{0} \otimes Q_{y}^{0} \quad + \sum_{\substack{y \in \cC_{n}\\ \dham(x_{T}, y_{T}) \leq \delta}} P_{y}^{1} \otimes Q_{y}^{1} \leq \Big( \sum_{\substack{y \in \cC_{n}\\ \dham(x_{S}, y_{S}) \leq \delta}} P_{y}^{0} \Big) \otimes \mathbb{1} \quad + \mathbb{1} \otimes \Big( \sum_{\substack{y \in \cC_{n}\\ \dham(x_{T}, y_{T}) \leq \delta}} Q_{y}^{1} \Big)\\
\leq \mathbb{1} \otimes \mathbb{1} \quad + \sum_{\substack{y \in \cC_{n}\\ \dham(x_{S}, y_{S}) \leq \delta}} P_{y}^{0} \otimes \sum_{\substack{z \in \cC_{n}\\ \dham(x_{T}, z_{T}) \leq \delta}} Q_{z}^{1},
\end{gather*}
where we applied the following operator inequality in the last step:
\begin{equation*}
A \otimes \mathbb{1} + \mathbb{1} \otimes B = \mathbb{1} \otimes \mathbb{1} + A \otimes B - (\mathbb{1} - A) \otimes (\mathbb{1} - B)\leq \mathbb{1} \otimes \mathbb{1} + A \otimes B,
\end{equation*}
for any $0 \leq A, B \leq \mathbb{1}$. Therefore,
\begin{equation*}
\Pi_{0}^{\theta} + \Pi_{1}^{\theta} \leq \mathbb{1} + \Pi_{c}^{\theta},
\end{equation*}
where $\Pi_{c}^{\theta}$ is the projector for the ``cross-game'', in which Bill has to unveil $0$ and Brian has to unveil $1$,
\begin{equation*}
\Pi_{c}^{\theta} = \sum_{x \in \cC_{n}} \ketbraq{x^{\theta}} \otimes \sum_{\substack{y \in \cC_{n}\\ \dham(x_{S}, y_{S}) \leq \delta}} P_{y}^{0} \otimes \sum_{\substack{z \in \cC_{n}\\ \dham(x_{T}, z_{T}) \leq \delta}} Q_{z}^{1}.
\end{equation*}
Clearly, $\tr \big( \rho [\Pi_{0}^{\theta} + \Pi_{1}^{\theta}] \big) \leq 1 + \tr \big( \rho \Pi_{c}^{\theta} \big)$, which combined with~\eqref{eq:p0-plus-p1} gives
\begin{equation*}
p_{0} + p_{1} \leq 1 + 2^{-n} \sum_{\theta \in \cC_{n}} \tr \big( \rho \Pi_{c}^{\theta} \big) = 1 + \tr \big( \rho \ave{\Pi_{c}^{\theta}} \big) \leq 1 + \norm{\ave{\Pi_{c}^{\theta}}},
\end{equation*}
where $\ave{\cdot}$ denotes averaging over $\theta$, i.e.~$\ave{\Pi_{c}^{\theta}} = 2^{- n} \sum_{\theta \in \cC_{n}} \Pi_{c}^{\theta}$, and $\norm{\cdot}$ denotes the Schatten $\infty$-norm. Changing the order of summation in $\Pi_{c}^{\theta}$ gives
\begin{equation*}
\Pi_{c}^{\theta} = \sum_{y, z \in \cC_{n}} \sum_{\substack{x \in \cC_{n}\\ \dham(x_{S}, y_{S}) \leq \delta\\ \dham(x_{T}, z_{T}) \leq \delta}} \ketbraq{x^{\theta}} \otimes P_{y}^{0} \otimes Q_{z}^{1}.
\end{equation*}
Now, it is clear that only the $x$-dependent part needs to be averaged:
\begin{equation*}
\ave{\Pi_{c}^{\theta}} = 2^{-n} \sum_{\theta \in \cC_{n}} \Pi_{c}^{\theta} = \sum_{y, z \in \cC_{n}} B_{yz} \otimes P_{y}^{0} \otimes Q_{z}^{1},
\end{equation*}
where
\begin{equation*}
B_{yz} = 2^{-n} \sum_{\theta \in \cC_{n}} \sum_{\substack{x \in \cC_{n}\\ \dham(x_{S}, y_{S}) \leq \delta\\ \dham(x_{T}, z_{T}) \leq \delta}} \ketbraq{x^{\theta}}.
\end{equation*}

The product $P_{y}^{0} \otimes Q_{z}^{1}$ yields orthogonal projectors so the norm of $\ave{\Pi_{c}^{\theta}}$ can be written as
\begin{equation}
\label{eq:cross-projector-norm}
\norm{\ave{\Pi_{c}^{\theta}}} = \max_{y, z \in \cC_{n}} \norm{B_{yz}}.
\end{equation}

To see which values of $y$ and $z$ maximize the norm we need to look more closely at the matrices $B_{yz}$. For every $\theta$ define $u(\theta) \in \cC_{n}$ to be the string that satisfies $[u(\theta)]_{S} = y_{S}$ and $[u(\theta)]_{T} = z_{T}$. Relabelling $x \to x \oplus u(\theta)$ yields
\begin{equation*}
B_{yz} = 2^{-n} \sum_{\theta \in \cC_{n}} \sum_{\substack{x \in \cC_{n}\\ \wham(x_{S}) \leq \delta\\ \wham(x_{T}) \leq \delta}} \ketbraq{(x \oplus u(\theta))^{\theta}}.
\end{equation*}
The constraints on the second sum can be relaxed by noting that $\wham(x_{S}) \leq \delta$ and $\wham(x_{T}) \leq \delta$ implies $\wham(x) \leq \delta$. Hence,
\begin{equation*}
B_{yz} \leq B_{yz}' = 2^{-n} \sum_{\theta \in \cC_{n}} \sum_{\substack{x \in \cC_{n}\\ \wham(x) \leq \delta}} \ketbraq{(x \oplus u(\theta))^{\theta}}.
\end{equation*}
Now the second sum is independent of $\theta$, hence, the summation over $\theta$ can be performed first. The product structure simplifies the summation
\begin{equation*}
\ketbraq{x^{\theta}} = \bigotimes_{k = 1}^{n} \ketbraq{\psi_{x_{k}}^{\theta_{k}}} \nbox{and} \sum_{\theta \in \cC_{n}} \ldots \iff \bigotimes_{k = 1}^{n} \sum_{\theta_{k} \in \{0, 1\}} \ldots,
\end{equation*}
which implies that
\begin{equation*}
2^{-n} \sum_{\theta \in \cC_{n}} \ketbraq{(x \oplus u(\theta))^{\theta}} = \bigotimes_{k = 1}^{n} \rho_{x_{k} \oplus y_{k}, x_{k} \oplus z_{k}}.
\end{equation*}
where
\begin{equation*}
\rho_{b, c} = \frac{1}{2} (\ketbraq{\psi_{b}^{0}} + \ketbraq{\psi_{c}^{1}}).
\end{equation*}
for $b, c \in \{0, 1\}$. Note that $\rho_{b, c} + \rho_{1 - b, 1 - c} = \mathbb{1}$ so they are diagonal in the same eigenbasis. Therefore, without loss of generality we can write
\begin{equation*}
\rho_{b, c} = \sum_{t \in \{0, 1\}} \lambda_{t}^{b \oplus c} \ketbraq{e_{t}^{b \oplus c}},
\end{equation*}
for $b, c \in \{0, 1\}$, where $\lambda_{0}^{b \oplus c} + \lambda_{1}^{b \oplus c} = 1$. In particular, we have
\begin{equation*}
\Big( \bigotimes_{k = 1}^{n} \rho_{x_{k} \oplus y_{k}, x_{k} \oplus z_{k}} \Big) \Big( \bigotimes_{k = 1}^{n} \ket{e_{v_{k}}^{y_{k} \oplus z_{k}}} \Big) = \bigotimes_{k = 1}^{n} \lambda_{x_{k} \oplus y_{k} \oplus v_{k}}^{y_{k} \oplus z_{k}} \ket{e_{v_{k}}^{y_{k} \oplus z_{k}}}.
\end{equation*}
Therefore, we also know the eigenbasis of
\begin{equation*}
B_{yz}' = \sum_{\substack{x \in \cC_{n}\\ \wham(x) \leq \delta}} \bigotimes_{k = 1}^{n} \rho_{x_{k} \oplus y_{k}, x_{k} \oplus z_{k}},
\end{equation*}
and the highest eigenvalue equals simply
\begin{equation*}
\norm{B_{yz}'} = \max_{v \in \cC_{n}} \sum_{\substack{x \in \cC_{n}\\ \wham(x) \leq \delta}} \prod_{k = 1}^{n} \lambda_{x_{k} \oplus y_{k} \oplus v_{k}}^{y_{k} \oplus z_{k}}.
\end{equation*}
Recall from~\eqref{eq:cross-projector-norm} that the expression we want to bound is 
\begin{equation*}
\max_{y, z \in \cC_{n}} \norm{B_{yz}'} = \max_{v, y, z \in \cC_{n}} \sum_{\substack{x \in \cC_{n}\\ \wham(x) \leq \delta}} \prod_{k = 1}^{n} \lambda_{x_{k} \oplus y_{k} \oplus v_{k}}^{y_{k} \oplus z_{k}} = \max_{a, b \in \cC_{n}} \sum_{\substack{x \in \cC_{n}\\ \wham(x) \leq \delta}} \prod_{k = 1}^{n} \lambda_{x_{k} \oplus b_{k}}^{a_{k}} .
\end{equation*}
Solving this maximization is easy (see Appendix~\ref{app:simple-maximization-problem} for details) and yields
\begin{equation*}
\max_{y, z \in \cC_{n}} \norm{B_{yz}'} = \sum_{\substack{x \in \cC_{n}\\ \wham(x) \leq \delta}} \prod_{k = 1}^{n} \lambda_{x_{k}} = \sum_{k = 0}^{ \lfloor \delta n \rfloor } {n \choose k} \lambda_{0}^{n - k} \lambda_{1}^{k},
\end{equation*}
where $\lambda_{0} = \max_{b, c} \lambda_{b}^{c}$ and $\lambda_{1} = 1 - \lambda_{0}$. Finally, as we know that
\begin{equation*}
\norm{\ave{\Pi_{c}^{\theta}}} = \max_{y, z \in \cC_{n}} \norm{B_{yz}} \leq \max_{y, z \in \cC_{n}} \norm{B_{yz}'}
\end{equation*}
we obtain the security guarantee of the form presented in~\eqref{eq:security-definition} with
\begin{equation*}
\varepsilon = \sum_{k = 0}^{ \lfloor \delta n \rfloor } {n \choose k} \lambda_{0}^{n - k} \lambda_{1}^{k}.
\end{equation*}
In the noiseless case ($\delta = 0$) we obtain $\varepsilon = \lambda_{0}^{n}$, while for $\delta > 0$ we use the Chernoff bound~\eqref{eq:chernoff-bound-below} for the binomial distribution to get
\begin{equation}
\label{eq:theory-final-result}
\varepsilon \leq \exp \bigg(- \frac{1}{2} \Big( \sqrt{\lambda_{1}} - \frac{\delta}{\sqrt{\lambda_{1}}} \Big)^{2} n \bigg),
\end{equation}
for any $\delta < \lambda_{1}$. It is also clear that the cross-game becomes easy for the dishonest parties if $\delta > \lambda_{1}$. Hence, $\lambda_{1}$ is the error threshold below which the cross-game becomes secure (asymptotically).

It is not clear whether the bit commitment game can be secure for any $\delta > \lambda_{1}$ but if that is the case it cannot be proven using the cross-game reduction.

Finally, we want to connect $\lambda_{1}$ to some measure of complementarity of the states emitted by Alice. We show in Appendix~\ref{app:explicit-diagonalisation} that
\begin{equation*}
\lambda_{1} = \frac{1 - t}{2},
\end{equation*}
where $t = \max_{b, c \in \{0, 1\}} \abs{\braket{\psi_{b}^{0}}{\psi_{c}^{1}}}$. For example for the perfect BB84 states we get $t = 1/\sqrt{2}$ and $\lambda = \lambda_{1} = 1/2 - 1/(2 \sqrt{2}) \approx 0.146$.

\subsubsection{Security for honest Bob}
Since Alice does not receive any information from Bob before the open phase, she remains fully ignorant about the value of his commitment.

\subsection{Fault-tolerant RBC with backreporting}
\label{sec:fault-tolerant-rbc-with-backreporting}

If Alice were to implement the protocol presented in Section~\ref{sec:fault-tolerant-rbc} using a Poisson-type source she would need to set the mean number of photons very high so that the honest Bob detects clicks in the majority of rounds. However, then the probability of multi-photon emissions becomes significant which allows dishonest Bob to cheat. Here, we present a modified protocol, in which Bob is required to backreport the rounds he missed, which closes this security loophole.

\subsubsection{The protocol}

The new protocol admits three parameters: the error allowance, $\delta$, the detection threshold, $\gamma$, and the mean number of photons per pulse emitted by Alice, $\mu$. We will see later that these paremeters determine the trade-off between robustness and security.

\printprotocol{1}

The security for honest Bob follows straightforwardly from Assumption 3. If the probability of detecting a photon does not depend on the basis chosen by Bob then there is no correlation between the valid set and Bob's measurement choice. Therefore, despite having received the valid set Alice remains fully ignorant about Bob's measurement basis.

The remainder of the paper investigates the trade-off between robustness and security (for honest Alice). First, we consider the asymptotic limit (the number of rounds goes to infinity) and we establish simple conditions that the hardware and protocol parameters must satisfy for the protocol to be both robust and secure. These conditions define a region in the space of parameters, which we call the \emph{feasible region}, and essentially tell us how good the equipment of the honest parties must be in order to implement the protocol securely. Finally, we show how to compute explicit security bounds for finite $n$.

\subsubsection{Asymptotic analysis}

In the asymptotic limit all the random variables are concentrated around their average values. Therefore, to see for what values of the parameters that the protocol is in principle feasible it is enough to calculate the relevant averages.

\paragraph*{Robustness.}

In the honest scenario if Bob has received at least one photon he will consider the round valid. For the protocol to go through he must have received at least $\gamma n$ non-vacuum rounds. Hence, what we care about in the asymptotic limit is that the expected number of non-vacuum rounds (as observed by Bob) exceeds the detection threshold, $\sum_{t = 1}^{\infty} \amsbb{E}[N_{r}] > \gamma n$. Expressing the expectation values through probabilities~\eqref{eq:nt-and-expectation-value} gives
\begin{equation*}
\sum_{t = 1}^{\infty} p_{r} > \gamma.
\end{equation*}
Replacing the sum by $1 - p_{0}$ and evaluating it using~\eqref{eq:pr-definition} we get
\begin{equation*}
e^{-\mu \eta} + \gamma < 1.
\end{equation*}
If the first condition is satisfied then we only need to ensure that Bob's error rate is below the threshold. As the probability of making an error (obtaining the wrong outcome despite measuring in the right basis) in each round equals $Q$, the expected Hamming distance between what Alice encoded and what Bob measured also equals $Q$. Therefore, the second condition for robustness is $
Q < \delta$.

\paragraph*{Security for honest Alice}
In the case of dishonest Bob we assume that his fibre and detectors are perfect ($\eta = 1$, $Q = 0$) and, also, let us make the following modification that gives (slightly) more power to Bob but also makes the analysis simpler: we will replace all the vacuum rounds by single-photon emissions. Then, $N_{m}  := \sum_{k = 2}^{\infty} N_{k}$, the number of multi-photon emissions, becomes the only relevant parameter. If Bob receives two (or more) photons he will measure the first photon in the $Z$ basis and the second one in the $X$ basis and he will obtain valid answers to both questions (namely opening $0$ and opening $1$). This means that whichever bit he decides to unveil his answer will always be accepted by Alice. Therefore, the multi-photon rounds Bob can ``win'' for free and such rounds do not contribute to security. Hence, dishonest Bob's optimal strategy will be to discard as many single-photon rounds as possible. It is clear that if Bob can discard all of them, he is left with multi-photon emissions only and no security is possible. Therefore, the first condition for security is: the number of multi-photon rounds is lower than Bob's detection threshold
\begin{equation*}
N_{m} < \lceil \gamma n \rceil.
\end{equation*}
After using up the entire backreporting allowance the number of valid rounds is $m = \lceil \gamma n \rceil$ but there are only $\lceil \gamma n \rceil - N_{m}$ (which is now guaranteed to be a positive number) single-photon rounds among them. Therefore, honest Alice thinks that they are playing a standard bit commitment game of $\lceil \gamma n \rceil$ rounds but there is a certain number of multi-photon ones, which Bob can win ``for free''. Hence, he can use his ``error allowance'' for the single-photon rounds and the security we achieve is that of playing a game of $\lceil \gamma n \rceil - N_{m}$ rounds with the non-fractional error allowance of $\delta \gamma n$, which gives the effective (fractional) error allowance of
\begin{equation*}
\delta' = \frac{\delta \gamma n}{\lceil \gamma n \rceil - N_{m}}.
\end{equation*}
In the proof (Section~\ref{sec:security-proof}) we find that we can only prove security if the effective (fractional) error allowance, $\delta'$, satisfies
\begin{equation*}
\delta' < \lambda_{1},
\end{equation*}
where $\lambda$ measures how complementary the bases used by Alice are. Hence, in our case we require
\begin{equation}
\label{eq:security-bound}
\delta \gamma n < (\lceil \gamma n \rceil - N_{m}) \lambda_{1}.
\end{equation}
In the asymptotic limit we look at the expectation value:
\begin{equation*}
\amsbb{E}[N_{m}] = [1 - e^{-\mu} (1 + \mu)] n,
\end{equation*}
which applied to~\eqref{eq:security-bound} gives
\begin{equation*}
e^{-\mu} (1 + \mu) + (1 - \frac{\delta}{\lambda_{1}}) \gamma > 1.
\end{equation*}

\paragraph*{Asymptotically feasible region.}
The asymptotically feasible region is defined as the region in the space of parameters that satisfies the robustness and security constraints, namely:
\begin{gather*}
e^{-\mu \eta} + \gamma < 1,\\
Q < \delta,\\
e^{-\mu} (1 + \mu) + (1 - \frac{\delta}{\lambda_{1}}) \gamma > 1.
\end{gather*}
It is clear that the first two inequalities (the robustness conditions) can be assumed to be equalities, which fixes $\delta$ and $\gamma$ and leaves us with only one, but fairly complicated, condition:
\begin{equation*}
e^{-\mu} (1 + \mu) + (1 - \frac{Q}{\lambda_{1}}) (1 - e^{-\mu \eta}) > 1.
\end{equation*}
This expression allows to verify whether for devices of given quality (quanitified by $\lambda_{1}$, $Q$ and $\eta$) there exists $\mu$ that makes the protocol both robust and secure.

\subsubsection{Finite-size effects}

The asymptotic analysis is relevant as $n \to \infty$ but in any practical scenario the number of rounds will be finite. Alice and Bob, having performed the protocol, will want explicit security guarantees (as a function of $n$). Note that robustness is verified experimentally, hence, there is no need to calculate it and security for honest Bob is perfect by assumption. Therefore, we are only left with security for honest Alice.

Let $E$ denote the event that Bob wins the cross-game. As shown in Section~\ref{sec:security-proof}, $\Pr[E]$ is an upper bound on the security parameter of the bit commitment protocol. Conditioning on the number of multi-photon emissions yields
\begin{equation}
\Pr[E] = \sum_{k = 0}^{n} \Pr[E | N_{m} = k] \Pr[N_{m} = k].
\end{equation}
Equation~\eqref{eq:theory-final-result} allows us to bound $\Pr[E | N_{m} = k]$ as long as the number of multi-photon emissions is below the threshold, $k_{t} = \gamma n (1 - \delta / \lambda_{1})$. For $k < k_{t}$ we have
\begin{equation}
\Pr[E | N_{m} = k] \leq \exp \bigg(- \frac{1}{2} \Big( \sqrt{(\gamma n - k) \lambda_{1}} - \frac{\delta \gamma n}{\sqrt{(\gamma n - k) \lambda_{1}}} \Big)^{2} \bigg),
\end{equation}
whereas for $k \geq k_{t}$ there is no security and the trivial bound, $\Pr[E | N_{m} = k] \leq 1$, is the best we can hope for. Noting that the random variable $N_{m}$ is Poisson-distributed with the parameter $p_{m} = 1 - e^{-\mu} (1 + \mu)$ allows us to evaluate the bound numerically.

\section{A simple maximization problem}
\label{app:simple-maximization-problem}

Suppose one is given two coins and is required to make $n$ independent coin flips. The goal is to maximize the probability of guessing correctly at least a certain fraction of the outcome string. Which coin should be used in every round and what is the optimal guess? Intuitively, it is clear that one should pick the more biased coin and always guess its more probable outcome. For completeness we provide a rigorous proof of this statement.

For $b, c \in \{0, 1\}$ let $\lambda_{b}^{c} \in [0, 1]$ be such that $\sum_{b} \lambda_{b}^{c} = 1$. In other words, $b$ corresponds to the outcome, while $c$ tells us which coin has been used. Let $y \in \cC_{n}$ be the string denoting which coin was used in each round and let $X$ be the random variable corresponding to the string of outcomes. Then for $x \in \cC_{n}$
\begin{equation*}
\Pr[X = x] = \prod_{k = 1}^{n} \lambda_{x_{k}}^{y_{k}}.
\end{equation*}
For an arbitrary string $z \in \cC_{n}$ the probability of being correct on at least $n (1 - \delta)$ position equals
\begin{equation*}
\Pr[\dham(X, z) \leq \delta] = \sum_{\substack{x \in \cC_{n}\\ \dham(x, z) \leq \delta}} \prod_{k = 1}^{n} \lambda_{x_{k}}^{y_{k}} = \sum_{\substack{x \in \cC_{n}\\ \wham(x) \leq \delta}} \prod_{k = 1}^{n} \lambda_{x_{k} \oplus z_{k}}^{y_{k}}.
\end{equation*}
Now, we want to see what values of $y, z \in \cC_{n}$ maximize this expression. Let us write down the summation corresponding to the $j$-th bit explicitly
\begin{equation*}
\sum_{\substack{x \in \cC_{n}\\ \wham(x) \leq \delta}} \prod_{k = 1}^{n} \lambda_{x_{k} \oplus z_{k}}^{y_{k}} = \lambda_{z_{j}}^{y_{j}} \Big[ \sum_{\substack{x \in \cC_{n - 1}\\ \wham(x) \leq n \delta/(n - 1)}} \prod_{k \neq j} \lambda_{x_{k} \oplus z_{k}}^{y_{k}} \Big] + (1 - \lambda_{z_{j}}^{y_{j}}) \Big[ \sum_{\substack{x \in \cC_{n - 1}\\ \wham(x) \leq (n \delta - 1)/(n - 1)}} \prod_{k \neq j} \lambda_{x_{k} \oplus z_{k}}^{y_{k}} \Big].
\end{equation*}
As the first square bracket is never smaller than the second one (because all the terms counted in the second bracket are also counted in the first one) we should always maximize $\lambda_{z_{j}}^{y_{j}}$. This observation is always true, regardless of the choices made for other rounds, and it applies to every round in the same way. Therefore, if $\lambda_{0} = \max_{b, c} \lambda_{b}^{c}$ and $\lambda_{1} = 1 - \lambda_{0}$ we find
\begin{equation*}
\max_{z \in \cC_{n}} \Pr[\dham(X, z) \leq \delta] = \sum_{\substack{v \in \cC_{n}\\ \wham(v) \leq \delta}} \prod_{k = 1}^{n} \lambda_{v_{k}} = \sum_{k = 0}^{ \lfloor \delta n \rfloor } {n \choose k} \lambda_{0}^{n - k} \lambda_{1}^{k},
\end{equation*}
which recovers the intuitively correct solution.

\section{Explicit diagonalisation of a mixture of two pure states}
\label{app:explicit-diagonalisation}

Let $\ket{\psi_{0}}$ and $\ket{\psi_{1}}$ be two pure states such that $\braket{\psi_{0}}{\psi_{1}} = c e^{i \phi}$. Then the eigenvalue decomposition of $\rho = \frac{1}{2} (\ketbraq{\psi_{0}} + \ketbraq{\psi_{1}})$ is
\begin{equation*}
\rho = \lambda_{+} \ketbraq{e_{+}} + \lambda_{-} \ketbraq{e_{-}},
\end{equation*}
where
\begin{gather*}
\lambda_{\pm} = \frac{1 \pm c}{2},\\
\ket{e_{\pm}} = \frac{1}{\sqrt{2 (1 + c)}} \Big[\ket{\psi_{0}} \pm e^{-i \phi} \ket{\psi_{1}} \Big].
\end{gather*}

\section{Chernoff bounds for the binomial distribution \cite{chernoff52}}

Let $X_{1}, X_{2}, \ldots, X_{n}$ be independent random variables taking on values 0 or 1. Then let $X = \sum_{i = 1}^{n} X_{i}$ and $\mu$ be the expectation value of $X$. Then for any $\delta > 0$ the following inequalities hold.
\begin{align*}
\Pr[X < (1 - \delta) \mu] &< \bigg(\frac{e^{- \delta}}{(1 - \delta)^{1 - \delta}} \bigg)^{\mu} \leq \exp \bigg(- \frac{\mu \delta^{2}}{2} \bigg), \nonumber\\
\Pr[X > (1 + \delta) \mu] &< \bigg(\frac{e^{\delta}}{(1 + \delta)^{1 + \delta}} \bigg)^{\mu} \leq \exp \bigg(- \frac{\mu \delta^{2}}{3} \bigg).
\end{align*}
The bounds can be expressed more conveniently as
\begin{align}
\label{eq:chernoff-bound-below}
\Pr[X < s] &< \exp \bigg(- \frac{1}{2} \Big( \sqrt{\mu} - \frac{s}{\sqrt{\mu}} \Big)^{2} \bigg),\\
\label{eq:chernoff-bound-above}
\Pr[X > s] &< \Big( \frac{\mu}{s} \Big)^{s} e^{s - \mu}.
\end{align}

\section{Remarks on Quantum Advantage} \label{quantumadvantage}

\subsection{Remarks 1}

Note that our protocol requires no communication from $A_1$ or $A_2$
at commitment or unveiling.  These agents need
trusted spacetime coordinates and reliable classical data storage, but 
do not need secure sites to keep data secret before or during the
protocol: Alice may keep the secret data about her quantum
states at some other secure site for later cross-checking.
No unconditionally secure classical bit commitment
protocol (relativistic or otherwise) allows this freedom, which
is potentially valuable in the technologically asymmetric case where Alice is less able
than Bob to maintain a network of secure sites.   
Conversely, if Alice shares the quantum state information securely with
$A_1$ and $A_2$ and they keep it securely, Bob's cheating possibilities are
quite constrained.
If Bob cheats at commitment, i.e. $B_1$ and $B_2$ send different bit
values for $b'$, he will be exposed as soon as $A_1$ and $A_2$ crosscheck, whether or not
he unveils.  If they send the same $b'$ and Bob unveils at any time
between $t_c$ and $t_c + (d/2c)$, but tries to cheat at this stage,
at least one of the $A_i$ will learn instantly --
without needing to cross-check -- since he must send inconsistent information about
$b$ to at least one of them.  Again, no unconditionally secure
classical bit commitment protocol has this last feature, 
which allows (in a relativistically appropriate sense) instant action
to be taken in the event of cheating at the unveiling stage.

\subsection{Remarks 2}

In the main part of our manuscript, we already briefly touched upon
different security models \cite{Kent2012b,Kaniewski2013a}. 
In particular, one can see \cite{Kaniewski2013a} our two site protocol as
achieving security in a setting where communication is not allowed
during certain phases of the protocol. For a bit commitment protocol,
one can thereby distinguish several natural phases, namely the commit
phase, the wait phase, and finally the open and verification phase.
One can then ask whether security is possible if communication is not
possible in any of these phases. 

This model is interesting because, while imposing minimum communication
constraints necessary to evade the impossibility
result~\cite[Sec.~III.A]{Kaniewski2013a}, it is also sufficiently
strong to allow for secure quantum bit commitment. One can show that
no classical protocol implemented in this model can be
secure~\cite[Sec.~IV.B]{Kaniewski2013a}.  From the above description
of the setup with multiple agents occupying different locations (see
Fig.~\ref{protocols}-b), we can readily deduce that the communication
restraints are satisfied. However, the setup enforces even stronger
communication constraints.
For an in-depth discussion and
illustrations concerning the different phases of bit commitment and
communication models we refer to~\cite{Kaniewski2013a}.

In~\cite{Kaniewski2013a} it was shown that there is a strict quantum
advantage when we take the minimum number of phases in which
communication is not allowed, namely the wait and open phase.
Communication during the open phase, however, was possible. As
mentioned before, the significant distance between Singapore and
Geneva naturally guarantees such a lack of communication and thus
enables secure bit commitment.

However, since of course the locations of our experimental labs are
fixed to be in Geneva and Singapore, the distance also guarantees a
lack of communication already during the commit phase itself. This
scenario, when communication is never allowed, has long been studied
in computer science~\cite{benor:bc}, where in particular it is known
that there exists a classical protocol for bit commitment that is
secure as long as communication is \emph{never} possible at all (see
e.g.~\cite{Kaniewski2013a} for further background).  Similar protocols
were derived using relativistic assumptions~\cite{Kent1999,Kent2005}, whose first
round is closely related to~\cite{benor:bc}.  
The classical protocol of~\cite{benor:bc} is secure against all 
quantum attacks~(\cite{Kent2005}; see also ~\cite{bcThesis}),
essentially because a successful quantum attack would imply
the possibility of communication between the separate sites.    
However, these classical protocols would not be
secure even against classical attacks if communication were possible during the commit phase, and so
the two site quantum protocol implemented here would offer a theoretical advantage in such a
setting.  

It is a very interesting open question, whether there exists an
natural experimentally implementable physical scenario in which 
communication during the commit phase is allowed, but
communication during the wait and open phase is forbidden.

Looking at the protocol
of~\cite{benor:bc} one could imagine that supplementing special
relativity with the assumption that only a finite amount of
communication is possible within a given time frame, and that an
infinitesimally small point in space can, by the holographic
priniciple, not store an arbitrary amount of randomness\,---\,required
to execute the protocol of~\cite{benor:bc}\,---\,can possibly lead to
such a scenario. However, these are only clues we can derive from a
particular classical protocol~\cite{benor:bc} and they do not by itself answer this interesting open question.

\section{Extension for three agents}
\label{three-agents}

Suppose that Bob sites his third agent, $B_0$, equidistant from $B_1$ and $B_2$ on the
shorter arc of a great circle joining them and preshare
$b$ and $r^{(b)}$ among all three agents, while Alice
locates her third agent, $A_0$, right next to $B_0$.  Then, instead of 
$B_1$ and $B_2$ committing the bit $a = b \oplus b'$, $B_0$
can do so by sending $b'$ to $A_0$ at $t_c$, while $B_1$ and $B_2$
unveil by sending $b$ and $r^{(b)}$ at $t_c + d/(2c)$. 
This allows $B_0$ to commit to a bit as soon as he learns it,
without presharing it with $B_1$ and $B_2$, while Alice is 
guaranteed that Bob's agents have collectively committed him 
from $t_c$ until $t_c + d/(2c)$. 

This freedom to commit bits in real time could be made more practical
if the $B_i$ preshare a large number of quantumly generated random bits $b_j$ and
commitment data $r^{(b_j )}$ and if $B_1$ and $B_2$ release
these in rapid sequence at coordinated times pre-agreed with $B_0$.
$B_0$ can then commit a bit $a= b_j \oplus b'$ at any time
of his choice, by sending $A_0$ both $b'$ and the value $j$
(chosen so that the unveiling times of $r^{(b_j)}$ guarantee
commitment for roughly $t_c + (d/ (2c))$.  
Adding a third site adds logistical complexity but no significant
technological challenge beyond those addressed in our implementation, which
thus demonstrates the practicality of secure real
time relativistic quantum commitment of bits acquired at a single location. 

\end{document}